\documentclass{iopart}

\usepackage{iopams}
\usepackage{graphicx}
\usepackage{color}
\usepackage{newlfont}
\usepackage{mathrsfs}
\usepackage{subfigure}
\usepackage{ulem}

\newcommand{\ket}[1]{|{#1}\rangle}

\newcommand{\sig}{\hat{\sigma}}
\newcommand{\sigz}{\hat{\sigma}_\mathrm{z}}
\newcommand{\sigx}{\hat{\sigma}_\mathrm{x}}
\newcommand{\sz}{\hat{s}_\mathrm{z}}
\newcommand{\sx}{\hat{s}_\mathrm{x}}
\newcommand{\sy}{\hat{s}_\mathrm{y}}
\newcommand{\s}{\hat{s}}

\begin{document}
	\title{Probing a composite spin-boson environment}
	\author{Neil P Oxtoby$^1$, \'Angel Rivas$^1$, Susana F Huelga$^1$, Rosario Fazio$^2$}
	\address{$^1$ Quantum Physics Group, STRI, School of Physics, Astronomy and Mathematics,
					 University of Hertfordshire, Hatfield, Herts AL10 9AB, United Kingdom}
	\address{$^2$ NEST CNR-INFM \& Scuola Normale Superiore, Piazza dei Cavalieri 7, I-56126 Pisa, Italy}
	\ead{\mailto{s.f.huelga@herts.ac.uk}}
	\submitto{\NJP}

	\begin{abstract}
		We consider non-interacting multi-qubit systems as controllable probes of an
        environment of defects/impurities modelled as a composite spin-boson environment.
      The spin-boson environment consists of a small number of quantum-coherent
      two-level fluctuators (TLFs) damped by independent bosonic baths.
      A master equation of the Lindblad form is derived for the probe-plus-TLF system.
        We discuss how correlation measurements in the probe system encode information
        about the environment structure and could be exploited to efficiently
        discriminate between different experimental preparation techniques, with
        particular focus on the quantum correlations (entanglement) that build up
        in the probe as a result of the TLF-mediated interaction.
      We also investigate the harmful effects of the composite spin-boson environment on initially prepared entangled
      bipartite qubit states of the probe and on entangling gate operations.
      Our results offer insights in the area of quantum computation using
      superconducting devices, where defects/impurities are believed to be a major
      source of decoherence.
	\end{abstract}

	\maketitle

\section{Introduction}
Superconducting qubits \cite{SCQreview} consist of
electronic nanocircuits embedding Josephson junctions whose dynamics
can, in certain parameter regimes, be restricted to  a
two-dimensional manifold. These qubits can be used as test-beds for
studying quantum mechanics at its most fundamental level \cite{LegJPCM02},
and are also potential candidates for the practical
implementation of quantum information processors
\cite{MakSchShnRMP01}.

An interesting and challenging aspect of
these endeavours is the process of decoherence, whereby qubits lose
their quantum-mechanical nature and are rendered dynamically equivalent
to their classical two-level counterparts.
In the case of a superconducting charge qubit, such as the
Cooper-pair box (CPB), dephasing (phase decoherence) is dominated by
low-frequency noise thought to be caused by interactions with
two-level fluctuators~\cite{NakPasTsaNAT99,ShnetalPRL05,NeeleyNATP08,LupARXIV08}
(TLFs) in the local environment. These TLFs may be charge traps
caused by defects/impurities in the Josephson junction or in the substrate.
There has been a substantial amount of research on TLFs causing
single qubit decoherence in Josephson junction systems.  Theoretical
works have concerned larger ensembles of TLFs, both incoherent
\cite{PalPRL02,GalPRL06,FaoPRL05,GriPRB05,SchNJP06,AbeMarPRB08,Faoro08} and
coherent \cite{ShnetalPRL05}, that randomly switch between two
configurations, producing low-frequency fluctuations in the relevant qubit parameters.
As discussed in detail in all these works, whenever a large number of
TLFs are very weakly coupled to the qubit, their effect can be described by a
conventional boson bath with a suitable chosen spectral density.
This is however not always the case --- there may be situations when
only one or a few impurities are important.
Indeed, Neeley, \etal \cite{NeeleyNATP08} have demonstrated the existence of
\textit{coherent} TLFs.
Lupa\c{s}cu, \etal \cite{LupARXIV08} provide evidence that these TLFs
are in fact genuine two-level systems.
In Neeley's experiment \cite{NeeleyNATP08} a single TLF that was
coupled to the qubit led to an avoided crossing in the qubit energy
spectrum.  The TLF was used as a proof-of-principle memory qubit, but
such TLFs will in general be detrimental to the operation of superconducting
qubits. It is therefore desirable to understand the behaviour of TLFs
in the vicinity of superconducting qubits in order to better-equip
quantum information scientists to manage the challenge of decoherence.

In view of its importance for understanding decoherence in superconducting
nanocircuits we study in this paper the effect that a few underdamped, coherent
TLFs have on the quantum dynamics of a charge qubit.
A single qubit coupled to a single coherent TLF, that was in turn
(under-)damped by a bosonic bath was already considered in~\cite{PalPRB08}.
We find this type of non-equilibrium environment~\cite{EmaPRA08}
to be an appealing model of the environment of a Josephson junction qubit because
TLFs may be damped by phonons in the substrate, for example.
We refer to this as a composite \textit{spin-boson environment}.
The case of a few TLFs, which will be considered here, is rather complex and
therefore we resort to extensive numerical calculations.
Our approach represents quite a general strategy for probing a composite
spin-boson environment, whereby we derive a Markovian master equation for
the average dynamics of the probe+TLFs after tracing out the external baths.
To allow for the random nature of TLF formation, we select the TLF
Hamiltonian parameters according to probability distributions
designated in \cite{ShnetalPRL05}. Of the many environmental
properties that could be studied in this scenario, we focus on
inferring the presence or absence of coherent coupling between the
TLFs.  This \textit{connectivity} of the TLFs has been identified as
important in the decoherence of a bipartite qubit system
\cite{YuanPRB08}, with further-reaching importance for quantum
computing.

Performing a more thorough treatment of this situation is increasingly
important in light of recent experiments \cite{NeeleyNATP08,LupARXIV08}
involving superconducting qubits coupling predominantly to only a small
number of TLFs.
In these experiments, measurements on the qubit were used to infer properties of the
TLF (although that wasn't the focus in \cite{NeeleyNATP08}).
That is, the qubit was used as a probe of the environment \cite{YaleSPIE03,AshNJP06}.
Probing properties of a small collection of TLFs, such as whether or not
they interact with each other, necessitates consideration of the full dynamics of the
probe plus TLFs under the influence of external baths.

From the theory point of view, identifying structural properties in
the environment could be done using some form of noise correlation
measurements. Here we focus on inferring
environmental features via the analysis of the entanglement that
will build up in a probe consisting of two non-(directly)
interacting qubits whose remote coupling is mediated by the TLFs 
present in the surroundings. When the probing is "local", so that
each probe qubit couples to just one single TLF, the absence of
quantum correlation generation would immediately signal a
non-connected environment, given that the probe qubits can only
become entangled if the fluctators would couple to each other.
Entanglement swapping in those circumstances has been discussed in
the literature \cite{rev1,rev2,rev3}. When probe qubits are subject
to the action of a \emph{few} TLFs, as it happens in qubit
realizations in the solid state, we will show that the remote
entanglement in the probe bears signatures that can be linked to the
connectivity in the environment and can in some cases be related to
monogamy constraints \cite{CofKunWooPRA00}. Given that bipartite entanglement has been shown
\cite{AudPleNJP2006} to be lower-bounded by combinations of
pseudo-spin observables, we also analyze what information can be
extracted from magnetization measurements along a given direction
(in the case considered here the magnetization along the z-direction
corresponds to the average charge) and study the power spectra of
magnetization observables using both single and bipartite probes. We
find that a double-qubit probe generally outperforms a single-qubit
probe, a result that could perhaps be expected given the extra
degrees of freedom available in the composite system. We supplement
our analysis of correlation measurements by investigating the
decoherence of composite probes initially prepared in a certain
maximally entangled state when subject to a composite spin-boson
environment, as well as the performance of entangling gate
operations when performed in the presence of this type of noise.

The paper is organized as follows. Section \ref{sec:system} sets the
scene by describing our model for the double-qubit probe and
spin-boson environment. A detailed derivation of the proposed master
equation as well as a discussion of its validity domain are presented in the appendix.
Numerical results for probing the connectivity of the spin-boson environment
are presented in section \ref{sec:connectivity}, using both probe entanglement
and estimated power-spectrum analyses.  We summarize and discuss
these results in section \ref{sec:discussion}, as well as compare the
double-qubit probe to a single-qubit probe.
In section \ref{sec:entdecay} we investigate the decoherence of
maximally-entangled Bell states induced by a composite spin-boson environment.
The performance of bipartite entangling gates in the presence of this form of noise is analyzed and discussed in section \ref{sec:gates}.
Section \ref{sec:conclusion} concludes.

\section{System}\label{sec:system}
\begin{figure}[!ht]
	\centering
	\includegraphics[width=0.5\textwidth]{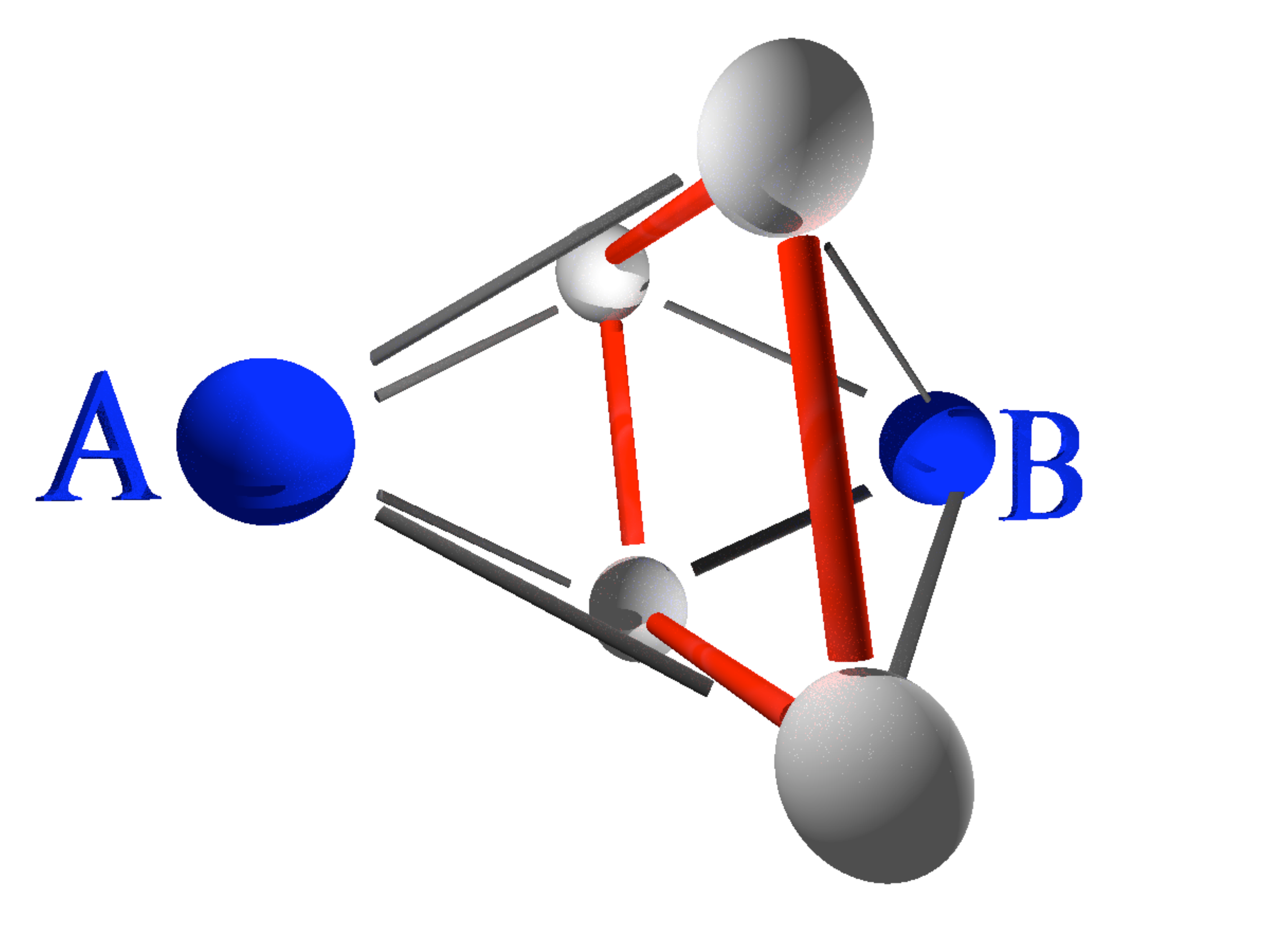}
	\caption{\protect\label{fig1}
	Double-qubit probe schematic.
    Blue spheres are the probe qubits, Alice and Bob.
    Grey spheres are the TLFs.
    Probe--TLF couplings ($\nu_j$) are depicted as black lines.
    TLF-TLF couplings ($\mu_{j,k}$) are specified by red lines.
    Each fluctuator is also subject to the action of a bosonic bath at a temperature $T$.
    Interactions between A and B are mediated by the TLFs, which lead to remote (TLF mediated)
entanglement generation in the probe.}
\end{figure}

The system we consider is illustrated in \fref{fig1}. It consists of two charge qubits (blue spheres)
acting as probes of an environment containing TLFs (grey spheres). Each qubit is coupled to a few
TLFs (black lines in the figure) but probe qubits are assumed to not directly couple to each other. In the numerical calculations we will
consider the case in which there are four TLFs.
Four TLFs is a balance between generating the desired spectral features
(requiring an ensemble of TLFs \cite{ShnetalPRL05,SchNJP06}) and
maintaining reasonable computation time (smaller Hilbert space).
Also, it may be the case that only a few TLFs will couple strongly
to a Josephson junction qubit, as in recent experiments
\cite{NeeleyNATP08,LupARXIV08}.
We should stress that the conclusions of our work do not depend on this choice.

In the charge basis, each qubit/TLF has a local free Hamiltonian
consisting of both longitudinal ($\sigz$) and transverse
($\sigx$) components:
$2\hat{H}_{\sigma} = \varepsilon \sigz + \Delta \sigx$.
In the eigenbasis the corresponding pseudo-spin
Hamiltonians are $2\hat{H}_\mathrm{s} = \Omega_\mathrm{s} \sz$,
where the spin frequency is
$\Omega_\mathrm{s}^2=\varepsilon^2+\Delta^2$.
(Throughout this article we denote Pauli operators in the
charge basis by $\hat{\sigma}$, and in the pseudo-spin basis by
$\hat{s}$.)  For simplicity, we ``engineer'' the probe qubit Hamiltonians
to have only longitudinal components ($\Delta_\mathrm{P}=0$).

    \paragraph*{Probe:}
    We label the identical probe qubits A and B.  Choosing uncoupled probe
    qubits for reasons that will become clear later, the total Hamiltonian
    for the probe is the sum
    $2\hat{H}_\mathrm{P} = \Omega_\mathrm{P} (\sz^\mathrm{A}+\sz^\mathrm{B})$.

    \paragraph*{Impurities:}
    We label the four TLF impurities with $j=1,2,3,4$.  The total Hamiltonian
    for the TLFs is then
    $2\hat{H}_\mathrm{TLF} =  \sum_{j=1}^4 \Omega_j \sz^{(j)} + \hat{V}_\mathrm{TLF}$,
    where $\hat{V}_\mathrm{TLF}$ describes coherent couplings between the TLFs,
    if they exist (defined below).
    Note that the pseudo-spin basis for the impurities is different to
    the probe (the $\sz^{(j)}$ axis is rotated relative to $\sz$)
    because the probe and TLF energies will differ, in general.
    Recent theoretical work~\cite{ShnetalPRL05} suggested specific
    distributions of these TLF energies in order to account for both
    low- and high-frequency noise observed in superconducting quantum systems.
    In our numerical study, we have adopted these distributions to determine
    the TLF bias energies $\varepsilon_j$ (linear distribution) and tunnel
    amplitudes $\Delta_j$ (log-uniform distribution).  Throughout the paper we
    will refer to $\Delta_j$ as the \textit{local field}. Again, note that our results are independent of the
    specific choice of frequency distribution and the same qualitative results can be derived when
    using a different functional form, e.g., a linear or a uniform distribution in a selected interval around the qubit
    frequency.

	\paragraph*{Interactions:}
    It is sensible to expect that the dominant interaction in a system of
    coherent two-level charges is an electrostatic one \cite{FaoPRL05,AstPRL04}.
    That is, charge-charge interactions.  We therefore assume bipartite ZZ
    interactions ($\sigz\otimes\sigz$) between subsystems.
    Within the TLFs we assume nearest-neighbour ZZ interactions
    of strength $\mu_{j,k}$, where $k = (j~\mathrm{mod}~4)+1$.
    We assume no direct interaction between the probe qubits A and B.
    So, $\hat{V}_{TLF}= \sum_{j=1}^4\mu_{j,k}\sigz^{(j)}\sigz^{(k)}$.
    For coupling strength $\nu_j$ between the $j$th impurity and the probe
    qubits, we have
    $\hat{V}_\mathrm{P} = \sum_{j=1}^4 \nu_j (\sigz^\mathrm{A} + \sigz^\mathrm{B}) \sigz^{(j)}$.
    We define $\hat{V}\equiv \hat{V}_\mathrm{P} + \hat{V}_\mathrm{TLF}$.
    In the numerical simulations, noting that we expect distant TLFs to have very little impact,
    we have assumed all couplings $\mu_{j,k}=\mu$ and all $\nu_j=\nu$ to avoid
    unnecessarily cumbersome results. Our conclusions are valid even when small variations in the parameters,
    of the order of $5-10 \%$, are considered.

\subsection{Master equation}\label{sec:mastereq}
The impurities are coupled to independent reservoirs of bosons (e.g., phonons
in the substrate), leading to dissipation (damping) as in the spin-boson
model \cite{RevModPhys.59.1}.
Under appropriate weak-coupling assumptions (see \ref{sec:MEderive} for our derivation),
the dissipative dynamics of the composite system (qubits plus damped TLFs) can be expressed
in the following Born-Markov master equation for
the joint state $\rho(t)$ of the probe plus impurities:
\begin{eqnarray}
    \dot{\rho}(t) &= -\rmi [\hat{H},\rho(t)]
                                     + \sum_j( \mathcal{D}_\mathrm{z}^{(j)}
                                                        +\mathcal{D}_+^{(j)}
                                                        +\mathcal{D}_-^{(j)})\rho(t) ,
    \label{eq:ME}
\end{eqnarray}
where the total Hamiltonian is
\begin{eqnarray}
    \hat{H} = \hat{H}_\mathrm{P} + \hat{H}_\mathrm{TLF} + \hat{V} .
    \label{eq:H}
\end{eqnarray}
The $\mathcal{D}$ superoperators \cite{WisMilPRA93b} represent decoherence in
the TLFs due to coupling with the bosonic baths, which are at temperature $T$.
The decoherence consists of dephasing
$\mathcal{D}_\mathrm{z}^{(j)}\rho = \Gamma_\mathrm{z}^{(j)}[\sz^{(j)}\rho\sz^{(j)} - \rho]$,
emission into the baths
$\mathcal{D}_{-}^{(j)}\rho = \Gamma_-^{(j)}[\s_-^{(j)}\rho\s_+^{(j)} -
(\s_+^{(j)}\s_-^{(j)}\rho + \rho\s_+^{(j)}\s_-^{(j)})/2]$,
and absorption from the baths
$\mathcal{D}_{+}^{(j)}\rho = \Gamma_+^{(j)}[\s_+^{(j)}\rho\s_-^{(j)} -
(\s_-^{(j)}\s_+^{(j)}\rho + \rho\s_-^{(j)}\s_+^{(j)})/2]$.
The TLF decoherence rates $\Gamma^{(j)}_{\mathrm{z},\pm}$ are proportional to
the respective dephasing, emission and absorption rates $\gamma_{\mathrm{z},\pm}$
(see \ref{sec:MEderive}), which are functions of the temperature and
spectral properties of the $j$th bosonic bath (with absorption dramatically
reduced at low temperatures).  Further, the TLF decoherence rates are also
functions of the ratio of local field to bias,
$\tan\theta_j \equiv \Delta_j/\varepsilon_j$ \cite{TsoetalNJP2007}.
To get a feel for the influence of this ratio, if we assume dissipation-limited
dephasing ($\gamma_\mathrm{z}=\gamma_-/2$) and sufficiently low temperature
($\gamma_+/\gamma_- \rightarrow 0$), then $\tan\theta_j$ dictates the dominance
of pure dephasing or relaxation in each TLF.  Specifically,
$\Gamma^{(j)}_\mathrm{z}/\Gamma^{(j)}_- = 1/\tan^2\theta_j$ so that pure dephasing
dominates the TLF decoherence for weak local fields, and relaxation dominates for
strong local fields.
Following \cite{ShnetalPRL05}, we distribute the random TLF parameters
$\varepsilon_j$, $\Delta_j$, and $\gamma^{(j)}_{\mathrm{z},\pm}$ as per
the distributions $P(\varepsilon_j) \propto \varepsilon_j$,
$P(\Delta_j) \propto 1/\Delta_j$,
$P(\gamma^{(j)}_{\mathrm{z},\pm}) \propto 1/\gamma^{(j)}_{\mathrm{z},\pm}$.
We choose these parameters to take values within the following moderate ranges:
$\varepsilon_j$ $\in$ $(1\pm0.5)\bar{\varepsilon}_j$;
$\Delta_j$ $\in$ $\bar{\Delta}_j \pm 0.5\min(\Omega_\mathrm{P},\bar{\Delta}_j) $;
$\gamma^{(j)}_{\mathrm{z},\pm}$ $\in$ $[\Omega_{\mathrm{min}}/6, \Omega_{\mathrm{min}}/2]$,
where $\Omega_\mathrm{min}$ is the minimum spin frequency amongst the TLFs.
We are free to select sensible values for the overbar quantities
$\bar{\varepsilon}_j$ and $\bar{\Delta}_j$, which we will reference to the tunable
probe frequency $\Omega_\mathrm{P}$.
Importantly, the TLFs are underdamped (therefore requiring a quantum-mechanical
description), so $\gamma^{(j)} < \Omega_j$.
We take the probe-TLF coupling to be uniform ($\nu_j=\nu$) and
weak compared with all of the TLF frequencies: $\nu=\Omega_\mathrm{min}/3$.
Our assumption of weak probe-TLF coupling simplifies the master
equation derivation significantly (see equation \ref{ApproxExp}; a detailed
discussion of an analogous situation can be found in \cite{CarmichaelSMQO1}),
and is in addition to the standard Born-Markov approximation of weak
TLF-bath coupling.
Weak probe-TLF coupling is in accord with the recent experiments of
\cite{NeeleyNATP08,LupARXIV08}, as well as the experiment of \cite{NakPasTsaNAT99}
(see \cite{GalPRL06}) where $\nu_j\sim$ MHz and $\Omega_j\sim\Omega_\mathrm{P}\sim$
GHz.
The TLF-TLF coupling is also assumed to be uniform $\mu_{j,k}=\mu$.
Note that the authors of \cite{ShnetalPRL05} point out that
$\nu$ must also be randomly distributed in order to realize $1/f$ noise in
the probe.
Generating the correct statistics is not within the scope of this paper
as we are interested in
describing effects when the environment is dominated by only a few fluctuators.

\subsection{Observable quantities}\label{sec:observables}
We restrict our knowledge to the probe subsystem (as would be the case in
an experiment).  A notable observable quantity on the probe is the
magnetization, which is related to a simple sum of Pauli operators:
$\hat{M}_\mathrm{x}(t)=\s_\mathrm{x}^\mathrm{A}+\s_\mathrm{x}^\mathrm{B}$.
The appeal of considering the probe magnetization is that it requires only
tractable, local measurements on each probe qubit.
That is, our results may be easily tested in an experiment.
It is worth noting that a result of Audenaert and Plenio \cite{AudPleNJP2006}
shows that measuring correlations $C_{xx/zz}=\langle \sig_{x/z} \otimes\sig_{x/z} \rangle$
along the XX and ZZ `directions' suffice to give a lower bound on the probe
entanglement.
This can remove the requirement for full tomographic (entanglement) measurements
when verifying or quantifying entanglement in the probe.

The time series resulting from measuring the probe's X-magnetization is
$M(t)=\langle{\hat{M}_\mathrm{x}(t)}\rangle$.
The mean-square power spectrum 
of $\hat{M}(t)$ is given by
\begin{eqnarray}
    S(\omega) &= \int_{-\infty}^{\infty} R(\tau)\rme^{-\rmi\omega \tau} \rmd\tau ,
    \label{eq:psd}
\end{eqnarray}
where the reduced auto-correlation function is
$R(\tau)\equiv\langle \hat{M}(t+\tau)\hat{M}(t)\rangle
 - \langle \hat{M}(t+\tau)\rangle\langle \hat{M}(t)\rangle$.
The angle brackets here denote the expectation value of an operator
$\langle \hat{x}\rangle = \Tr[\hat{x}\rho(t)]$.
In reality, $M(t)$ is a discrete quantity (the data is a time series), and
so the power spectrum obtained is an estimate (the Fourier transform of the reduced
autocorrelation of the time series $M(t)$).  This estimate of average power
as a function of frequency can theoretically be improved (by increasing the duration
of the experiment, for example), but this may not be practical in reality.
Strictly speaking, $M(t)$ must be a wide-sense stationary process (time-independent
first and second moments) for the power spectrum to exist.

\section{Results: Detecting the presence of coupling between the TLFs}
\label{sec:connectivity}
Can probe observables reveal the degree of connectivity of a composite spin-boson
environment?  In this section we present numerical results showing that
this is indeed the case. Observable quantities we consider are the probe magnetization,
its estimated power spectrum, and the remote entanglement between the probe qubits.
Entanglement generated between probe qubits that are initially in a separable
state is primarily due to the structure of the spin-boson environment (e.g.,
the presence or absence of TLF-TLF interactions in the surroundings of the
probe).

Initially we set the probe to be in a state orthogonal to the
$\sz$ eigen-axis:
$\ket{\psi_\mathrm{P}(0)} = \ket{+}_\mathrm{A}\ket{+}_\mathrm{B}$.
The TLFs are assumed to be initially in a zero-temperature thermal state,
i.e., the ground state of $\hat{H}_\mathrm{TLF}$, which we denote as $\ket{g}$.
So, $\ket{\phi_\mathrm{TLF}(0)} = \ket{g}$.
We plot observable quantities as a function of the ratio of TLF-TLF
coupling strength to probe-TLF coupling strength, $\mu/\nu$, which is often
believed to be small (see \cite{AshNJP06}, for example).

\Sref{sec:PS} considers the estimated power spectrum of $\langle\hat{M}_\mathrm{x}(t)\rangle$.
\Sref{sec:entprobe} considers the build up of entanglement in the probe.

\subsection{Power spectrum of $\langle \hat{M}_\mathrm{x} \rangle$}\label{sec:PS}
	\subsubsection{TLFs with weak local fields.}
\begin{figure}[!ht]
{\centering
	\subfigure[~$\varepsilon_\mathrm{P} = 10 \bar{\varepsilon}_j$.]%
		{\label{fig2a}\includegraphics[width=0.5\textwidth]{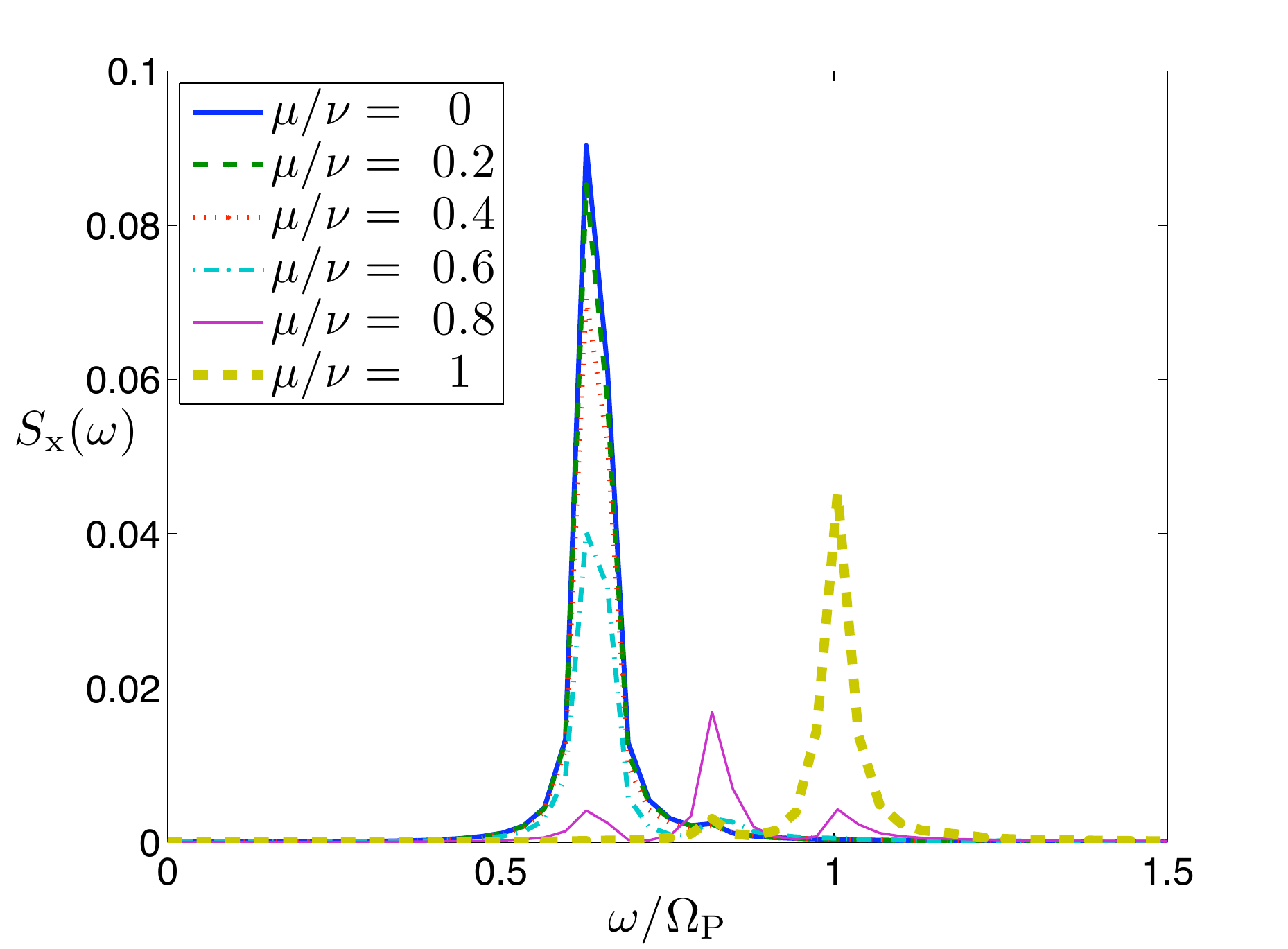}}%
	\subfigure[~$\varepsilon_\mathrm{P} = 3\bar{\varepsilon}_j$.]%
		{\label{fig2b}\includegraphics[width=0.5\textwidth]{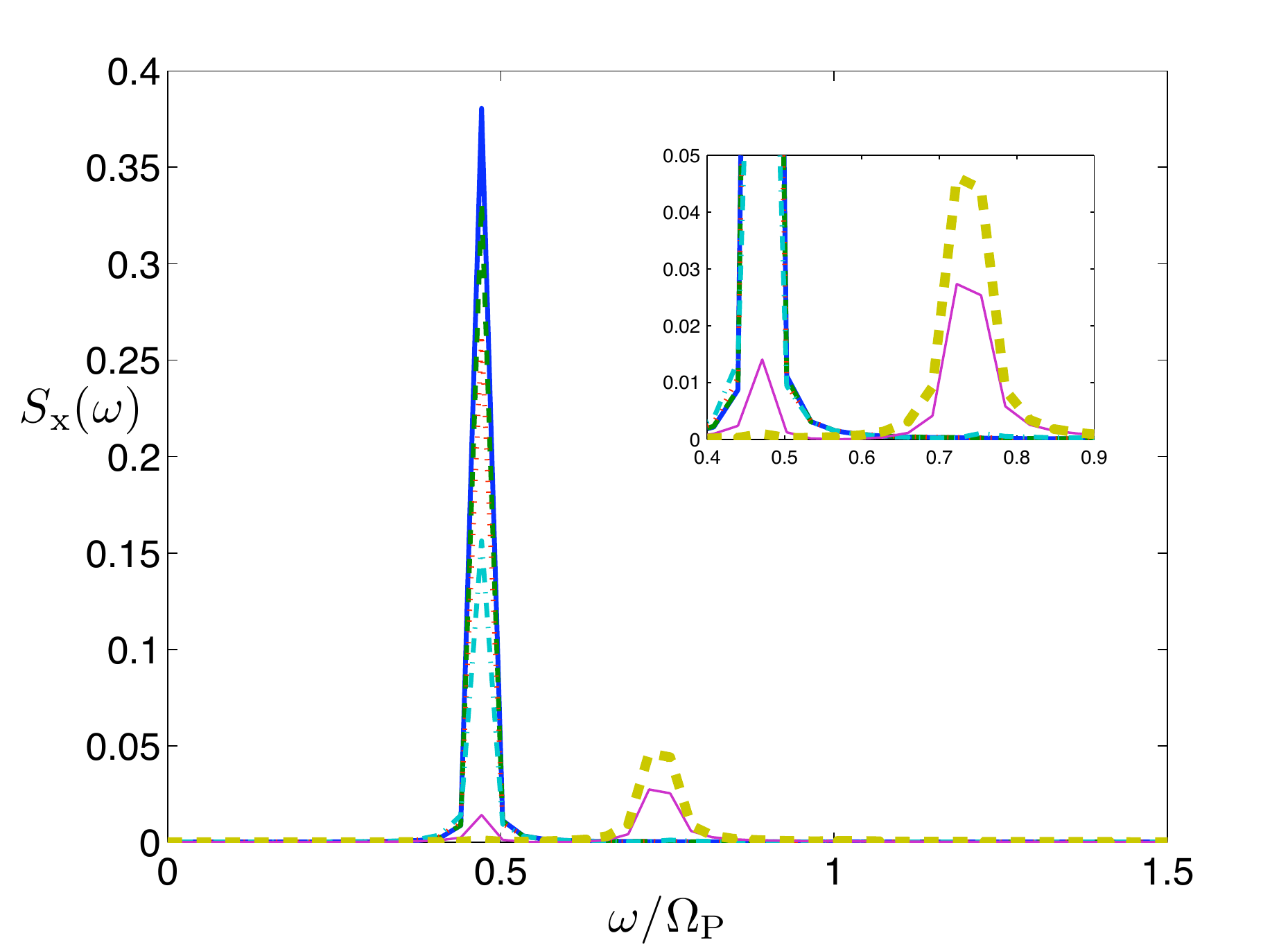}}
	\subfigure[~$\varepsilon_\mathrm{P} = \bar{\varepsilon}_j$.]%
		{\label{fig2c}\includegraphics[width=0.5\textwidth]{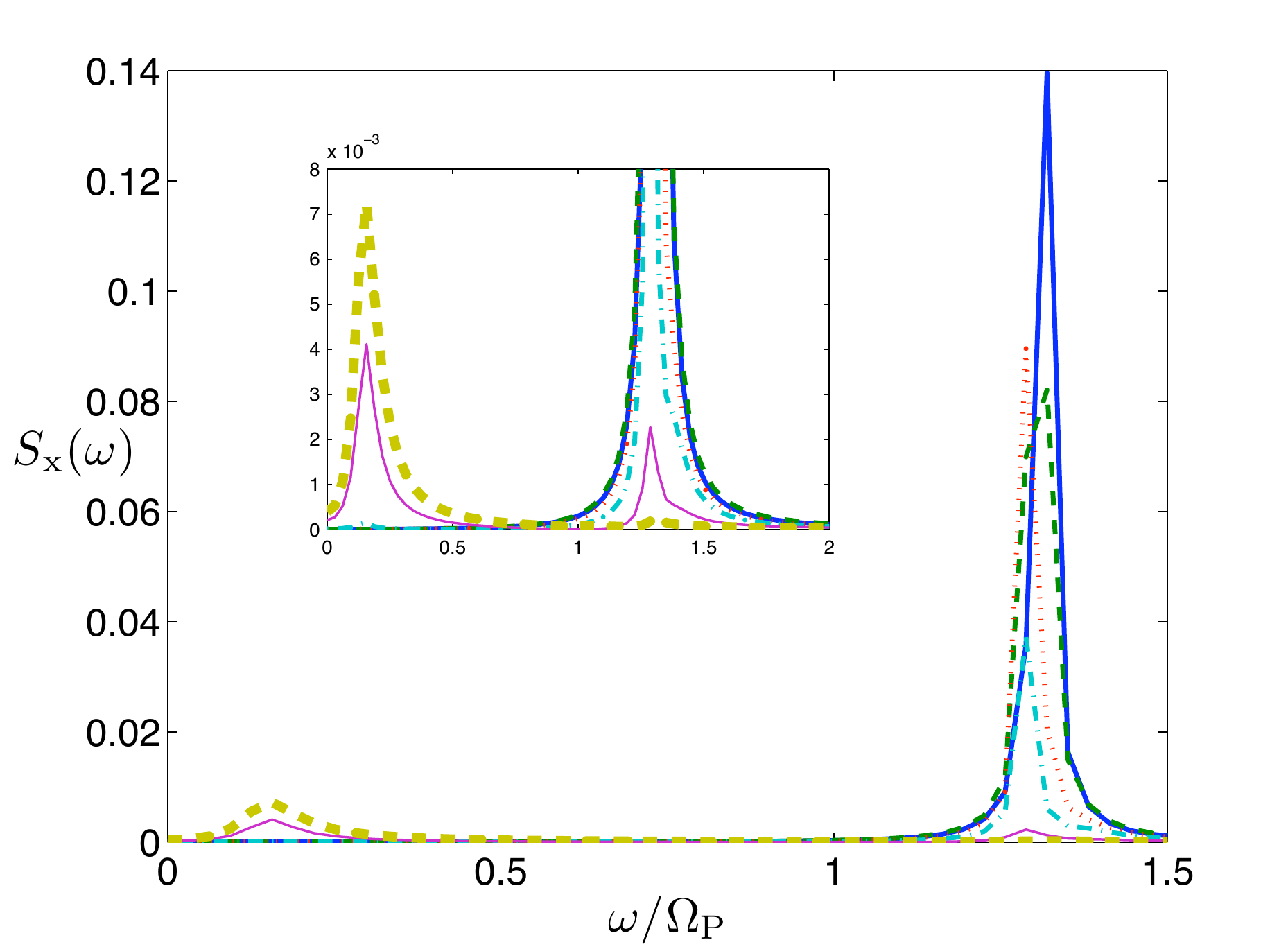}}%
	\subfigure[~Control spectrum: $\nu=0$.]%
		{\label{fig2d}\includegraphics[width=0.5\textwidth]{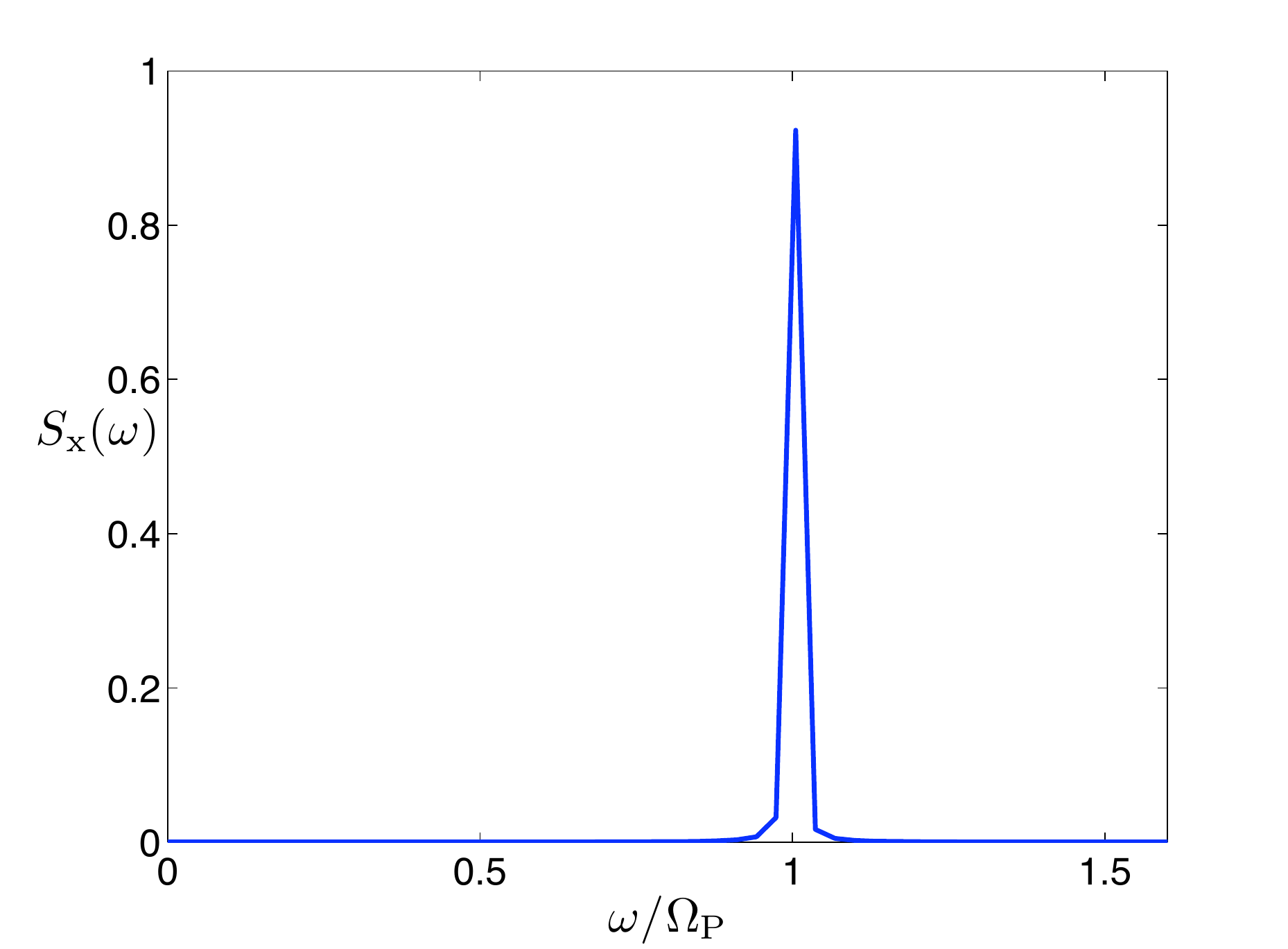}}
	\caption{\protect\label{fig2}
	Estimated power spectrum of $\langle\hat{M}_x(t)\rangle$ for weak local
	fields $\tan\bar{\theta}_j = 1/3$ with the initial probe state
	$\ket{\psi(0)} = \ket{++}$.
	Sampling parameters (units of $\Omega_\mathrm{P}=\varepsilon_\mathrm{P}$):
	$\delta f=1/200$, $t_s=0.05$, $f_N=10$.
	Figure \protect\ref{fig2d} is the power spectrum for an isolated probe.
	The insets show the smaller peaks magnified.
	The relative heights and positions of the peaks can be used to distinguish
	unconnected TLFs ($\mu=0$, thick solid blue line) from highly connected
	TLFs ($\mu\rightarrow\nu$, thick dashed gold line).
	}
	}
\end{figure}

	Consider the case of ``weak'' local fields in the spin-boson environment where
    $\tan\theta_j \sim 1/3$ 
    (specifically $\bar{\Delta}_j/\bar{\varepsilon}_j = 1/3$,
    where TLF dephasing dominates relaxation: $\Gamma_\mathrm{z}/\Gamma_-\sim 10$).
    In our probing of TLF connectivity, we tune the ratio of probe--TLF splitting
    $\varepsilon_\mathrm{P}/\bar{\varepsilon}_j$ to 10, 3 and 1, corresponding to
    figures \ref{fig2a}, \ref{fig2b} and \ref{fig2c},
    respectively.  We observe two effects:
    1. For weaker TLF interconnectivity $\mu/\nu \lesssim 0.6$,
    a decrease is observed in the height of the single dominant peak in the spectrum;
    2. For stronger TLF interconnectivity $\mu/\nu\approx1$, the spectrum splits into
    multiple peaks, the most dominant of which is shifted in frequency relative to
    $\mu/\nu=0$.  This is visible in \fref{fig2} where the gold traces
    ($\mu/\nu=1$) are the most qualitatively different from the blue traces ($\mu=0$).
    The power in the signal $\langle\hat{M}_\mathrm{x}(t)\rangle$ redistributes from one
    dominant frequency for unconnected TLFs ($\mu=0$), to multiple frequencies as
    $\mu/\nu$ approaches unity (highly connected TLFs).
    Remarkably, the most dominant peak for highly connected TLFs $\mu/\nu=1$ is
    qualitatively similar to the case of an isolated probe (\fref{fig2d})
    --- a single peak at $\omega=\Omega_\mathrm{P}$ --- although the peak visibility
    (height) is noticeably less than the isolated probe.

	\subsubsection{Effect of TLF local field.}  
	Above we have established that we can distinguish between highly connected
    TLFs ($\mu=\nu$) and unconnected TLFs ($\mu=0$) for weak local field strength
    $\Delta_j < \varepsilon_j$.
    It is interesting to ask how stronger local fields affect our
    ability to distinguish these two values of $\mu$.
    To explore this, we increase the ratio of TLF field strength to splitting:
    $\tan\bar{\theta}_j = \bar{\Delta}_j/\bar{\varepsilon}_j$,
    which was less than 1 in \fref{fig2}.  This varies the range
    for $\Delta_j$, which (we remind the reader) we have taken to be
    $\bar{\Delta}_j \pm 0.5 \min(\bar{\Delta}_j,\Omega_\mathrm{P})$.  This range
    ensures a sensible variation in the TLF local field strengths of no greater
    than one-half of the probe frequency. (We assume that the TLF energies are
    distributed over a relatively small range as might be expected for systematically 
		formed impurities/defects.)

\begin{figure}[ht]
	\centering
	\subfigure[~$\tan\bar{\theta}_j = 1/3$.]%
		{\label{fig3a}%
		\begin{minipage}{0.45\textwidth}
			\includegraphics[width=\textwidth]{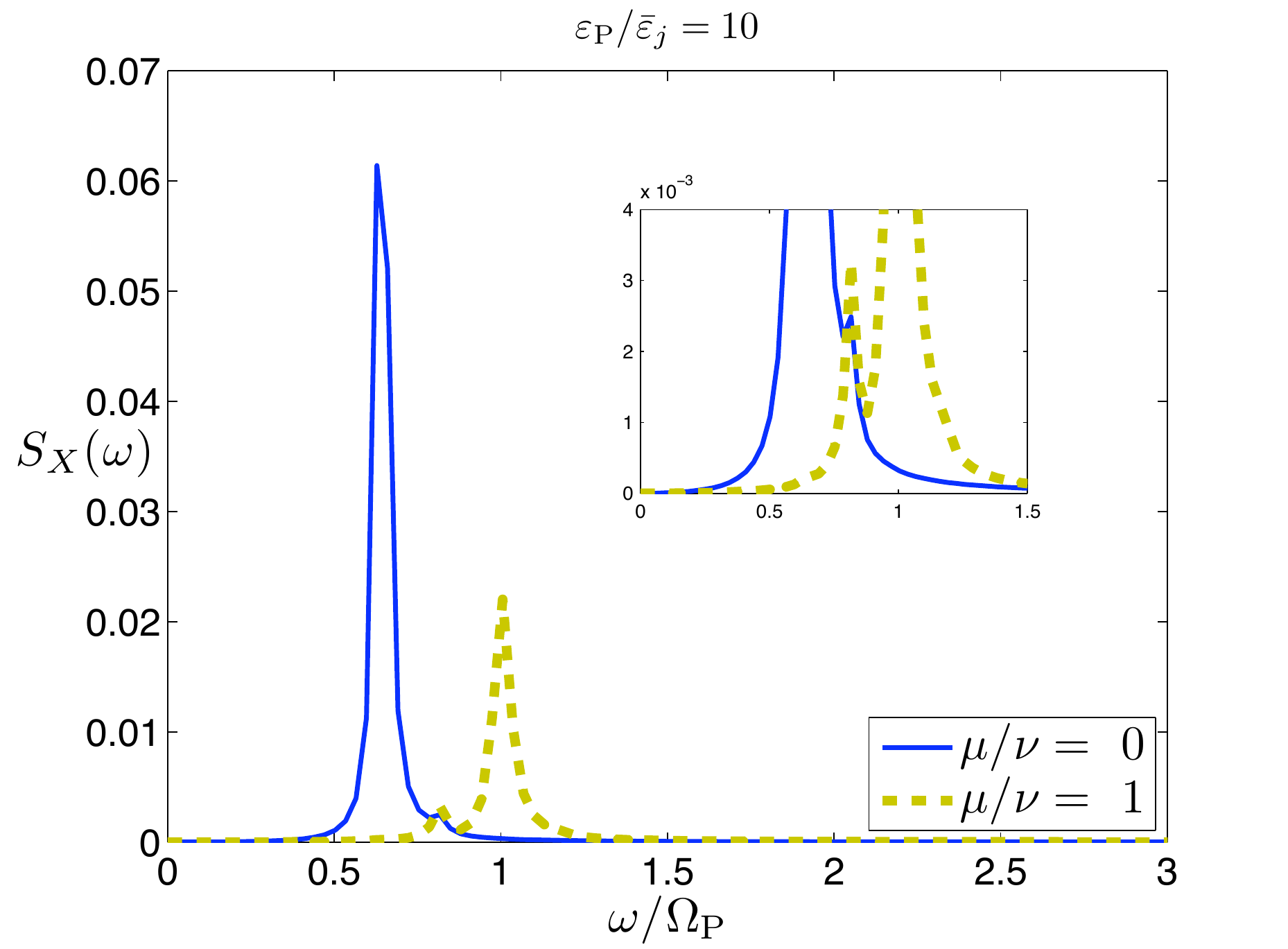}\\
			\includegraphics[width=\textwidth]{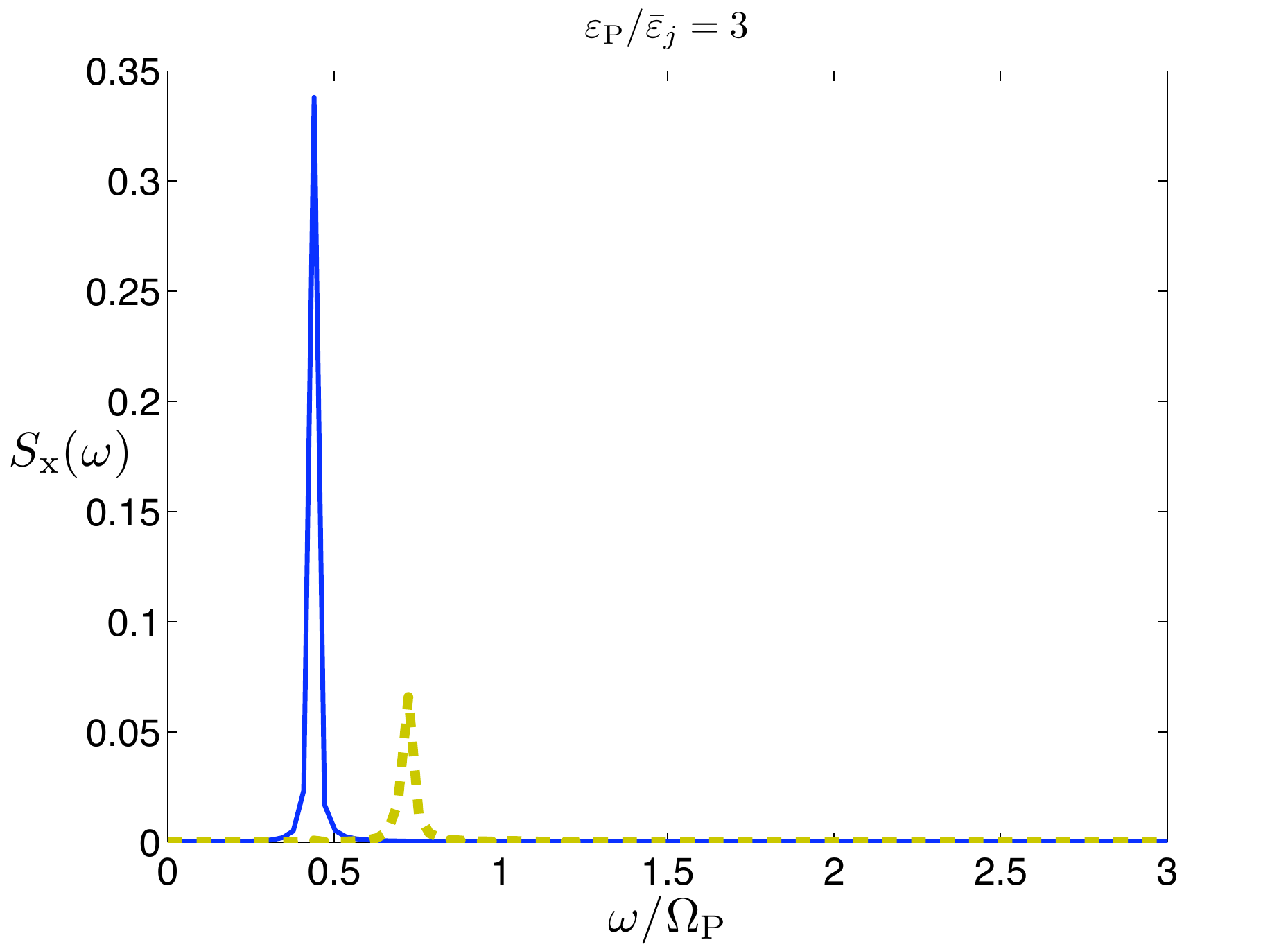}\\
			\includegraphics[width=\textwidth]{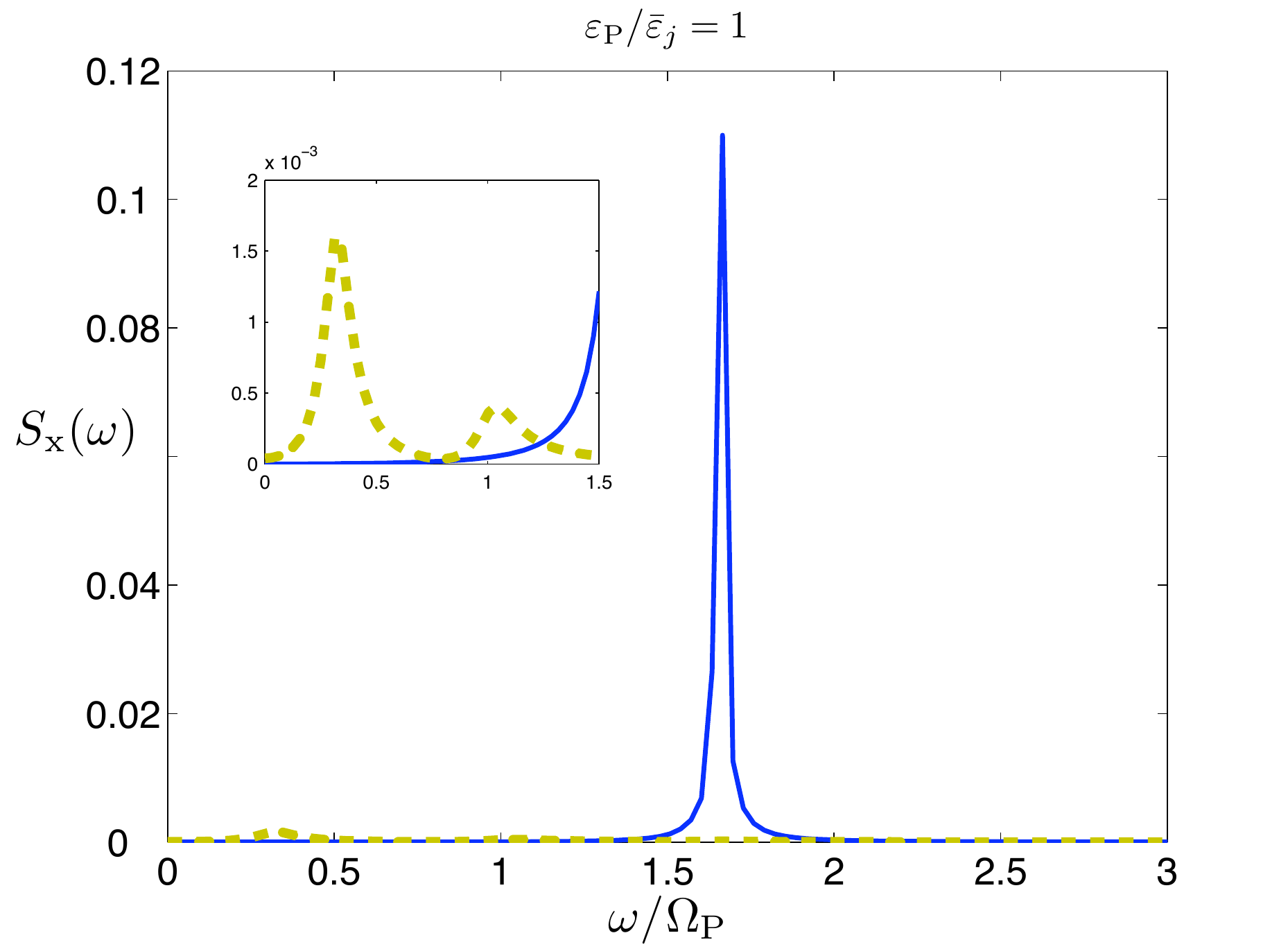}%
		\end{minipage}}%
	\subfigure[~$\tan\bar{\theta}_j = 3$.]%
		{\label{fig3b}%
		\begin{minipage}{0.45\textwidth}
			\includegraphics[width=\textwidth]{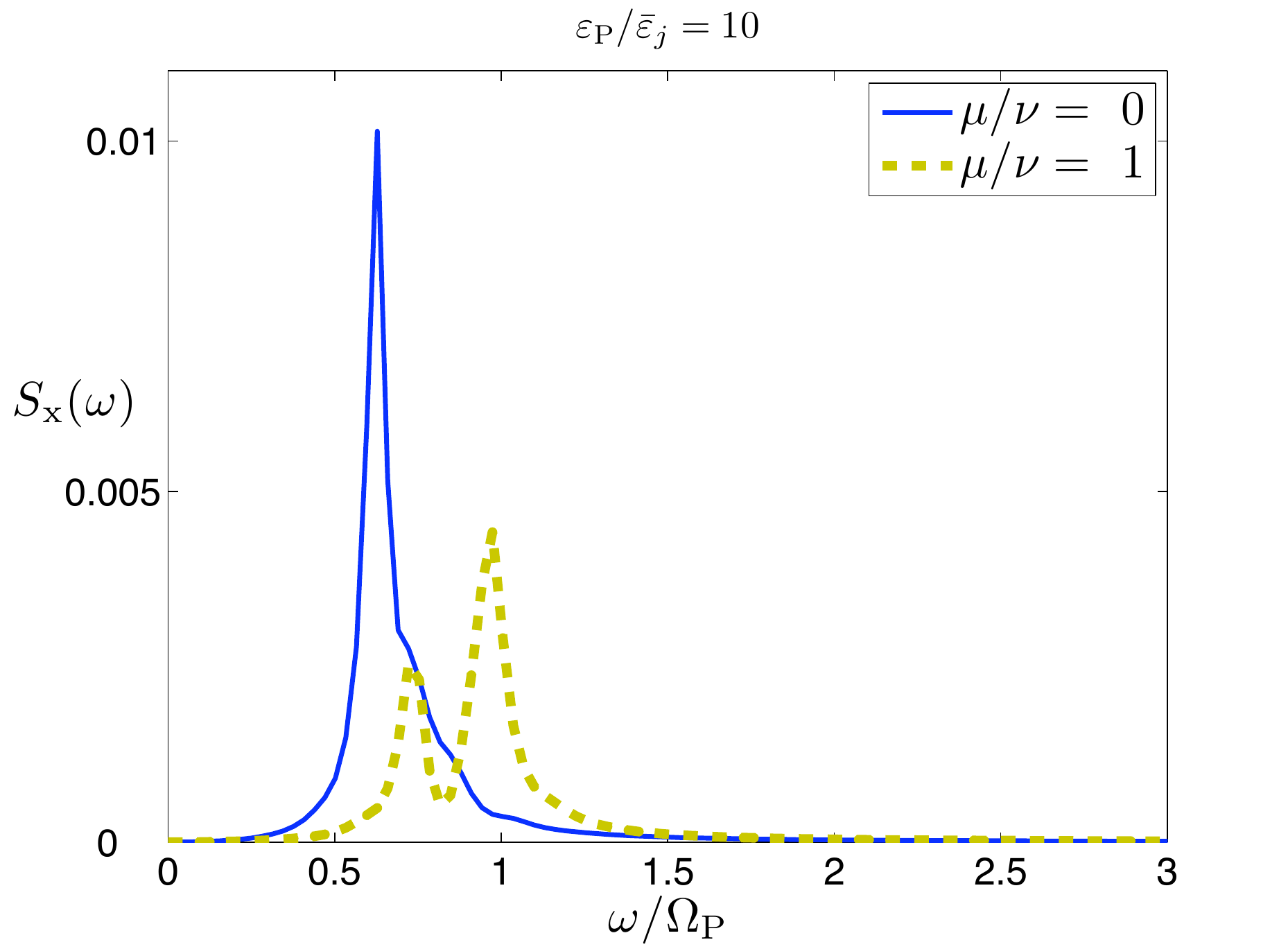}\\
			\includegraphics[width=\textwidth]{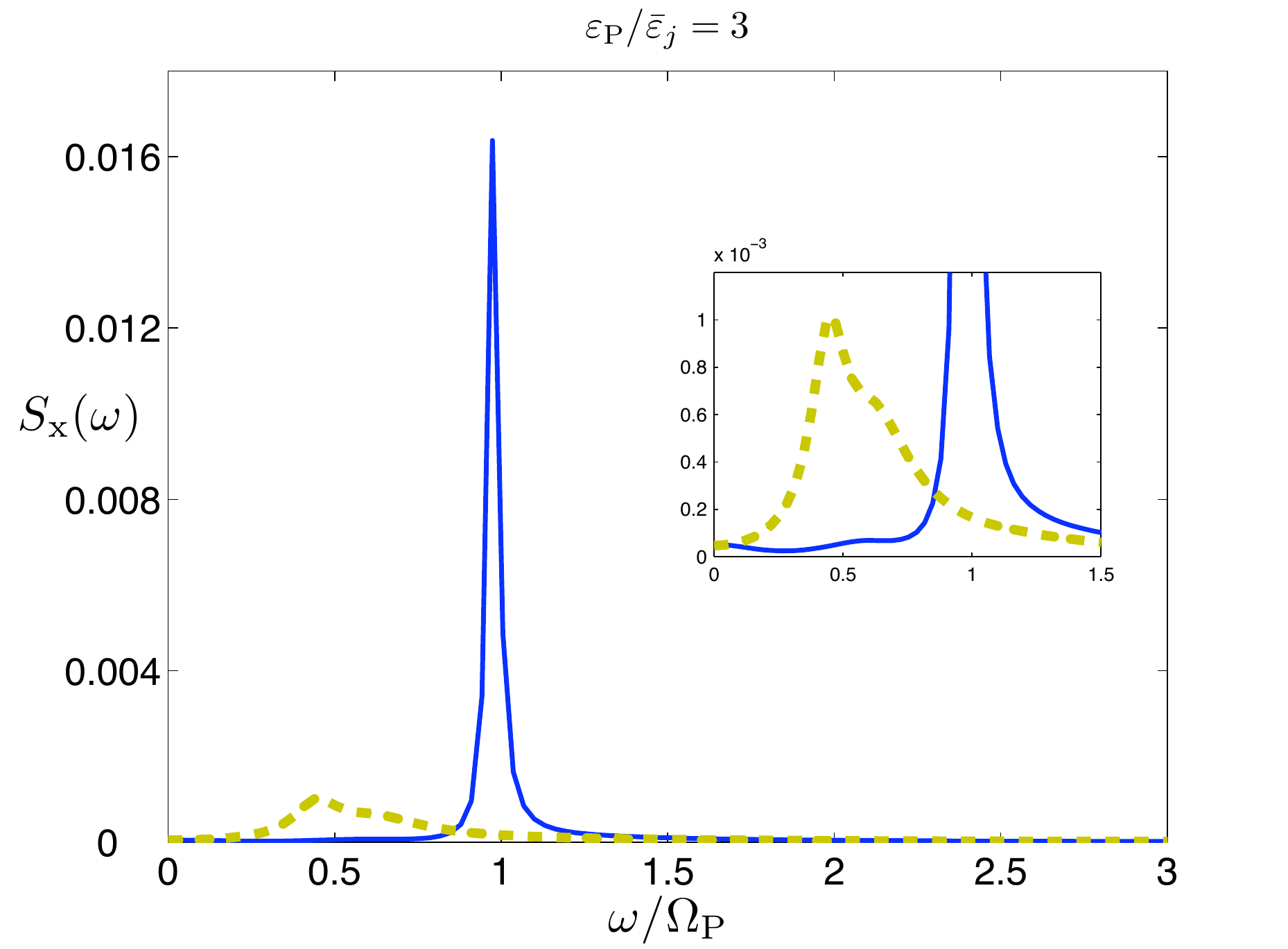}\\
			\includegraphics[width=\textwidth]{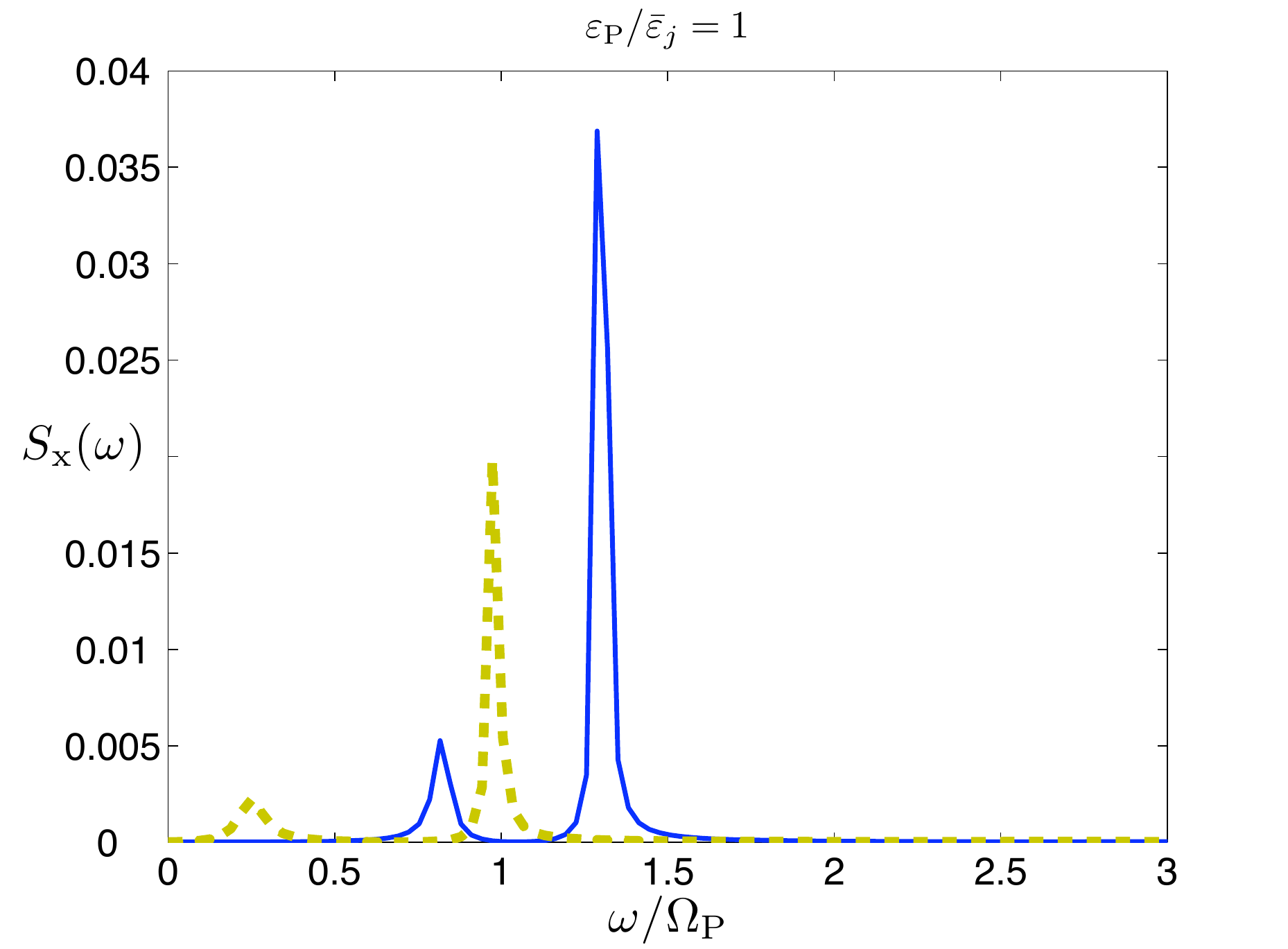}%
		\end{minipage}}
	\caption{\protect\label{fig3}
	Estimated power spectrum of $\langle\hat{M}_x(t)\rangle$ for
	$\varepsilon_\mathrm{P}/\bar{\varepsilon}_j = 10$ (top row),
	$3$ (second row) and $1$ (third row). 
	As the local field strength $\tan\theta_j=\Delta_j/\varepsilon_j$ increases
	from \protect\ref{fig3a}--\protect\ref{fig3b}, TLF relaxation
	dominates TLF dephasing. See text for discussion.
	}
\end{figure}

    As the local field increases $\tan\bar{\theta}_j > 1$,
    two different effects occur:
    1. Larger local fields cause relaxation to dominate over dephasing in the
    TLF decoherence; 2. The TLF eigenstates increasingly align towards the
    $\sigx$ axis, and seem to have a decreasing effect on the probe, perhaps
    because the interaction is of the $\sigz\otimes\sigz$ type.
    Evidence to support this is shown in \fref{fig3}, which
    shows a peak visibility (height) reduction
    of about one order of magnitude as $\tan\bar{\theta}_j$ is increased by
    one order of magnitude from 1/3 to 3 [\ref{fig3a} to
    \ref{fig3b}].  So, although extra features appear in the
    power spectrum, they become increasingly difficult to observe.
    Despite this, there is at least one plot in each column for which $\mu=0$
    and $\mu=\nu$ are distinguishable.  Thus it is apparent that tuning the
    probe frequency (selecting a row in figure \ref{fig3})
    allows these TLF connectivities to be distinguished for a wide
    range of values of the TLF local field strength (we obtained similar results
    for $\tan\bar{\theta}_j=1$).

\subsection{Probe entanglement}\label{sec:entprobe}
Interactions between probe qubits are mediated by the TLFs and entanglement between the probe qubits can be generated in this indirect way.
We now consider using the probe entanglement to distinguish between
connected and unconnected TLFs.  We use the logarithmic negativity as a
measure of bipartite entanglement between the probe qubits, defined
as \cite{PhysRevA.65.032314}
\begin{equation}
    E_\mathrm{P} \equiv \log_2||\rho_\mathrm{P}^{T_\mathrm{A}}||_1 ,
\label{eq:logneg}
\end{equation}
where $||\cdot||_1$ denotes the trace norm, and $\rho^{T_\mathrm{A}}$
is the partial transpose of $\rho$.

	\subsubsection{TLFs with weak local fields.}
\begin{figure}[!ht]
{\centering
	\subfigure[~$\varepsilon_\mathrm{P} = 10\bar{\varepsilon}_j$.]%
		{\label{fig4a}\includegraphics[width=0.33\textwidth]{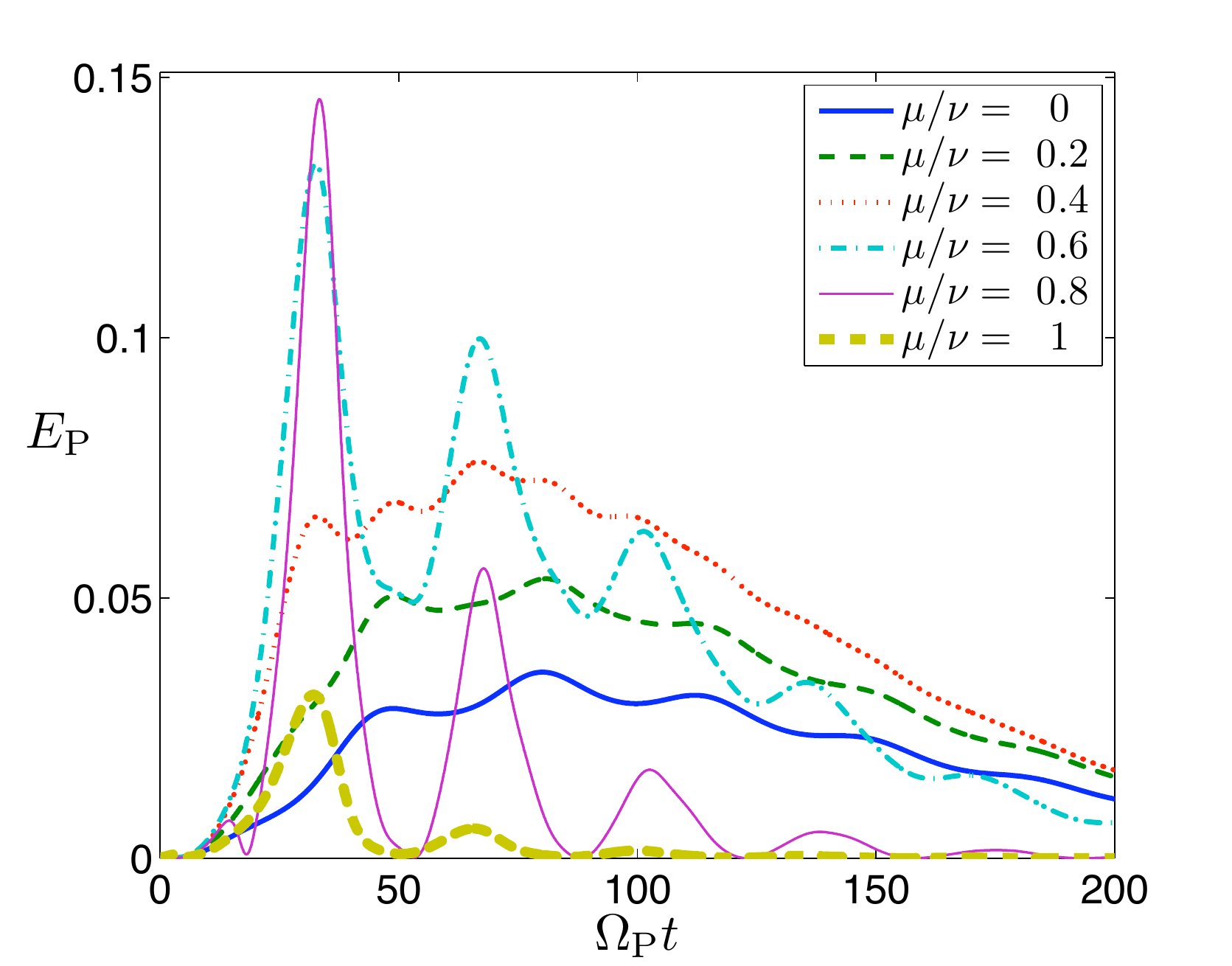}}%
	\subfigure[~$\varepsilon_\mathrm{P} =  3\bar{\varepsilon}_j$.]%
		{\label{fig4b}\includegraphics[width=0.33\textwidth]{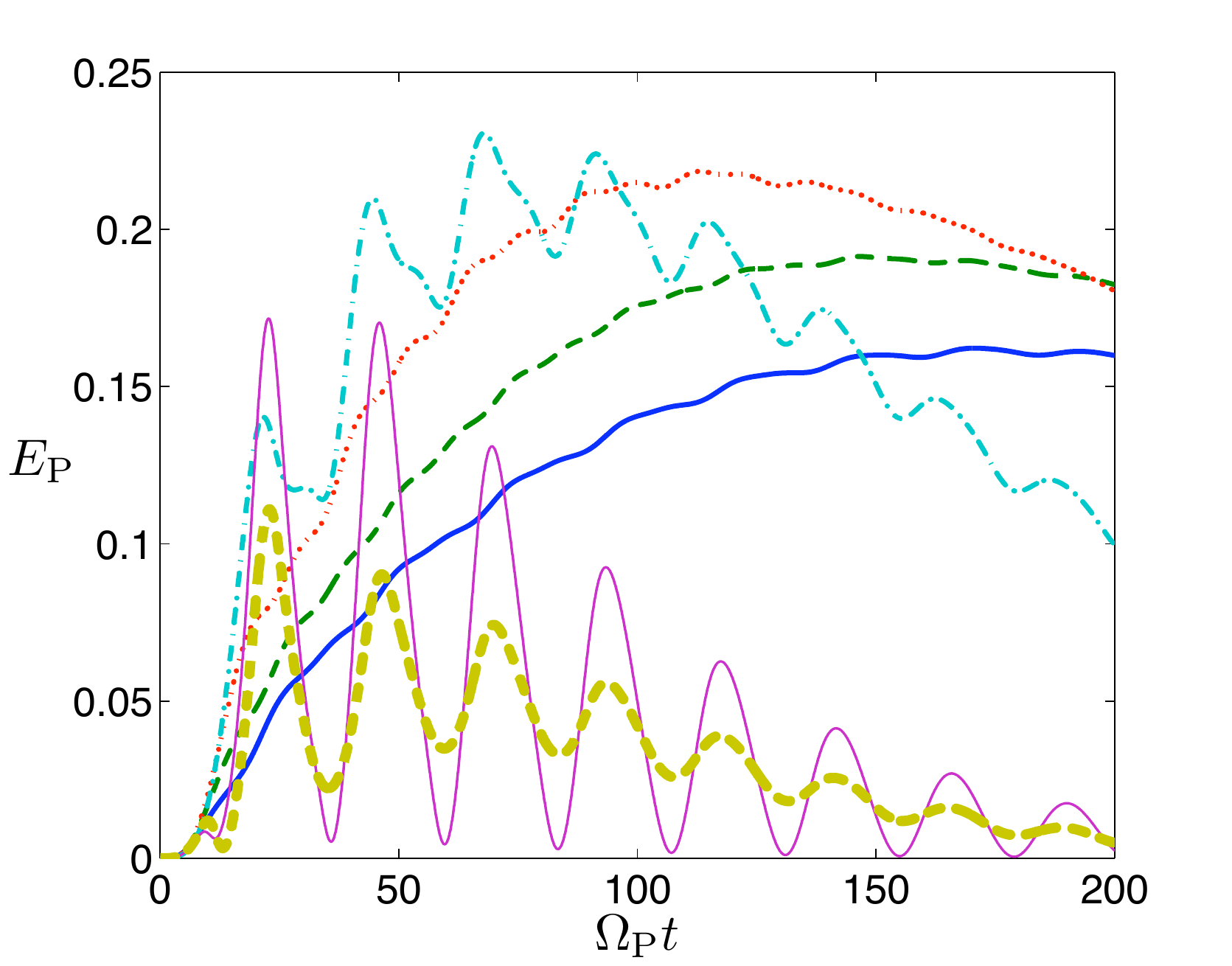}}%
	\subfigure[~$\varepsilon_\mathrm{P} =   \bar{\varepsilon}_j$.]%
		{\label{fig4c}\includegraphics[width=0.33\textwidth]{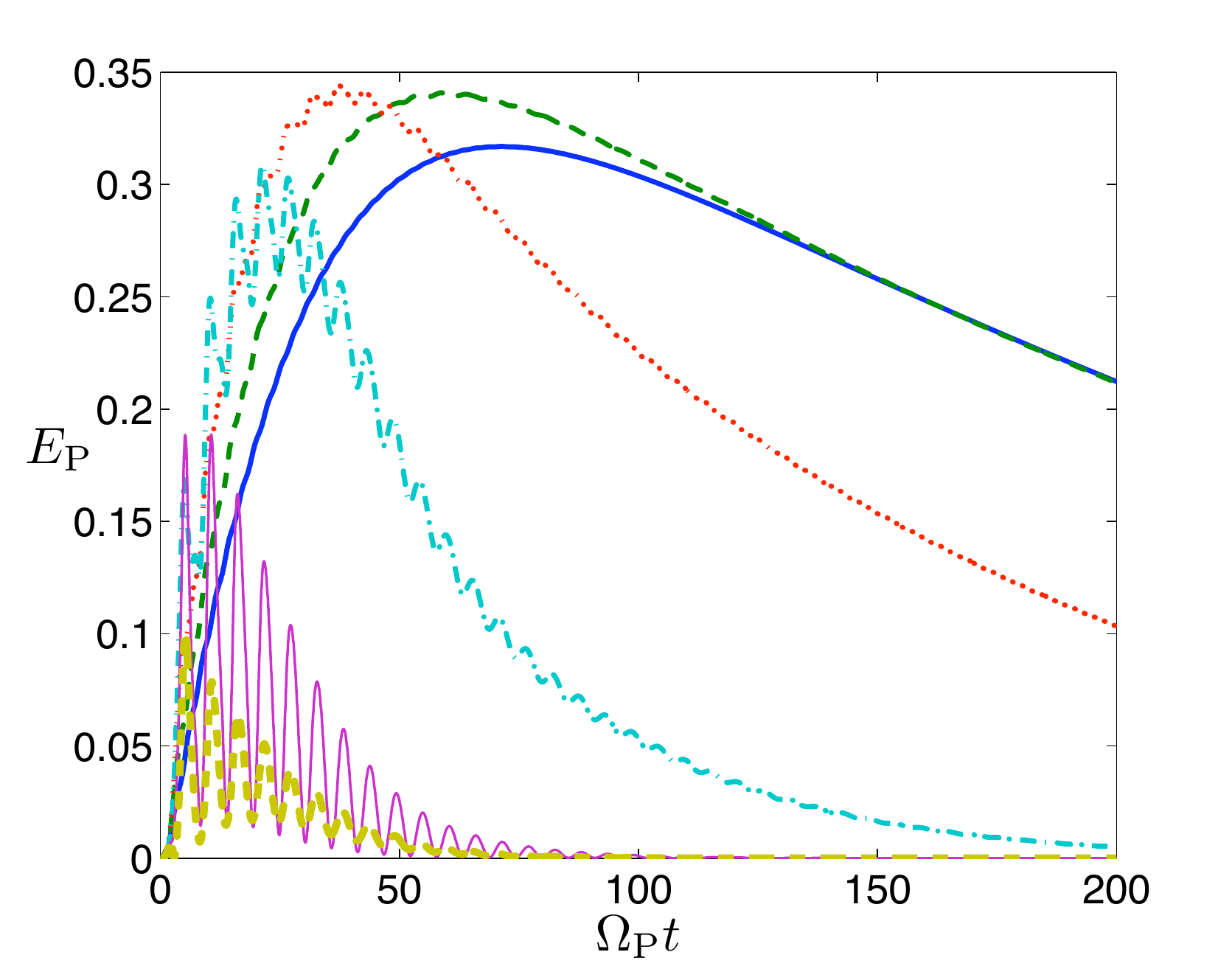}}\\
	\caption{\protect\label{fig4}
	Entanglement (logarithmic negativity) in the probe as a function of probe
	cycles, for small TLF local field
	strength $\tan\bar{\theta}_j = 1/3$.
	Note the distinct difference between strongly interacting TLFs ($\mu\approx\nu$)
	and non-interacting TLFs ($\mu=0$).}
}
\end{figure}

	Figure \ref{fig4} shows the logarithmic negativity of $\rho_\mathrm{P}(t)$
    for the same data sets as in \fref{fig2} (weak local fields
    $\tan\bar{\theta}_j = 1/3$).  We make two observations.
    Firstly, tuning the probe frequency provides only one benefit for distinguishing
    $\mu=0$ from $\mu\neq0$.  This is evidenced by the remarkable qualitative
    similarity between figures \ref{fig4a}, \ref{fig4b}, and
    \ref{fig4c}.  The benefit is the number of probe qubit cycles required
    to distinguish the two cases (time is shown in units of probe qubit cycles).
    Secondly, it is clear that strongly coupled TLFs (the gold line) cause less
    entanglement to generate within the probe. This can be explained by invoking the
    concept of entanglement monogamy \cite{CofKunWooPRA00,DawetalPRA05}.
    As the TLF-TLF connectivity increases, the indirect
    link between the two probe qubits is weakened, and so remote entanglement generation slows.
It is important to emphasize which partitions one should consider when invoking monogamy arguments. The relevant quantity is the entanglement shared between each probe qubit and a given TLF; this is the quantity that is sensitive to entanglement sharing in two ways:
(1) It decreases as the entanglement in the probe builds up, independently of
the connectivity in the TLF environment;
(2) It ``feels'' the TLF-TLF coupling in the sense that, within a selected time interval,
the stronger the fluctuators couple, the smaller the entanglement between probe qubit and TLF becomes.
As a result, remote entanglement builds up more slowly when fluctuators couple so that it quickly
degrades in a decohering environment, as illustrated by results in \fref{fig5}.

	\subsubsection{Effect of the TLF local field.}
\begin{figure}[ht]
	\centering
	\subfigure[~$\tan\bar{\theta}_j = 1/3$.]%
		{\label{fig5a}%
		\begin{minipage}{0.45\textwidth}
			\includegraphics[width=\textwidth]{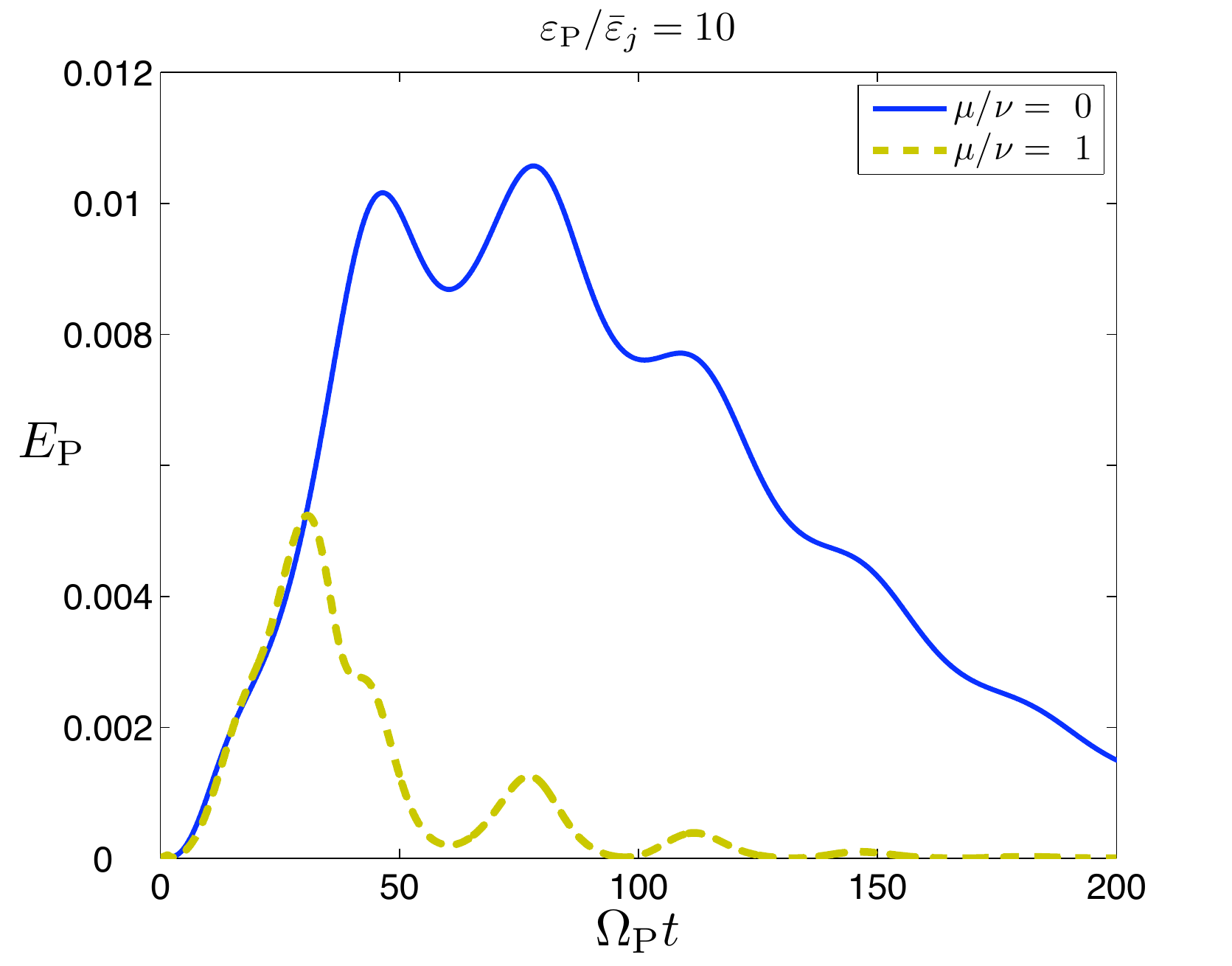}\\
			\includegraphics[width=\textwidth]{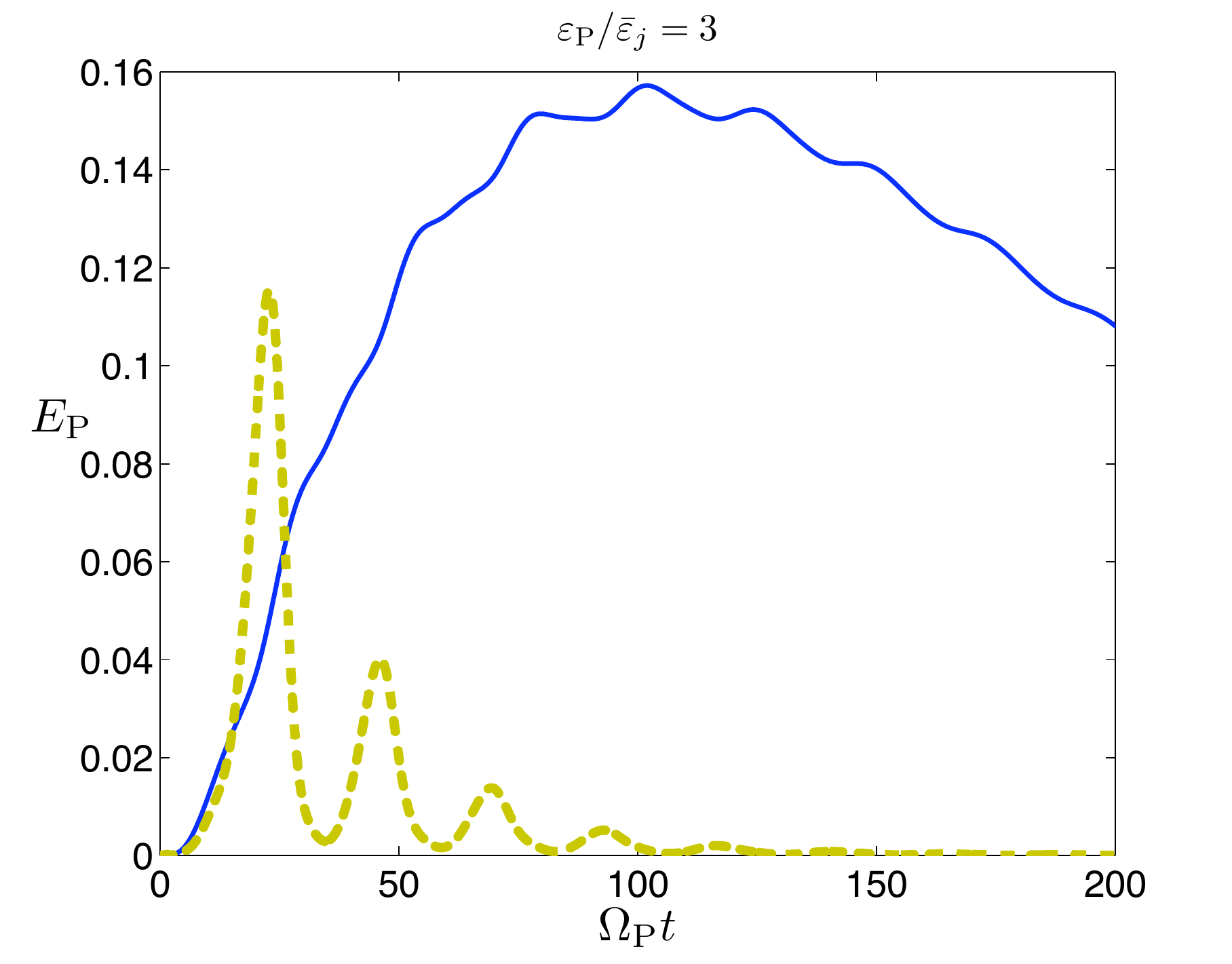}\\
			\includegraphics[width=\textwidth]{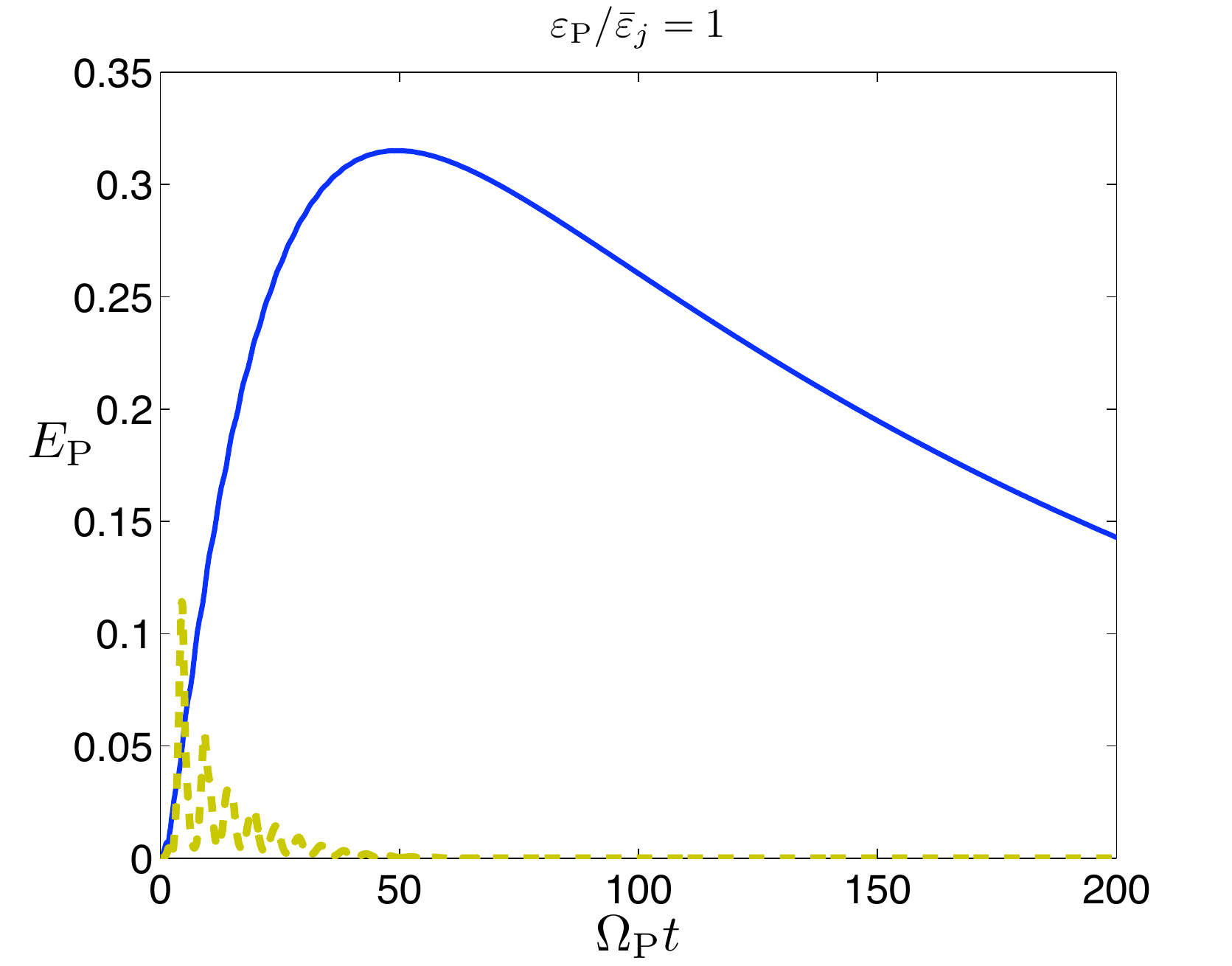}%
		\end{minipage}}%
	\subfigure[~$\tan\bar{\theta}_j = 3$.]%
		{\label{fig5b}%
		\begin{minipage}{0.45\textwidth}
			\includegraphics[width=\textwidth]{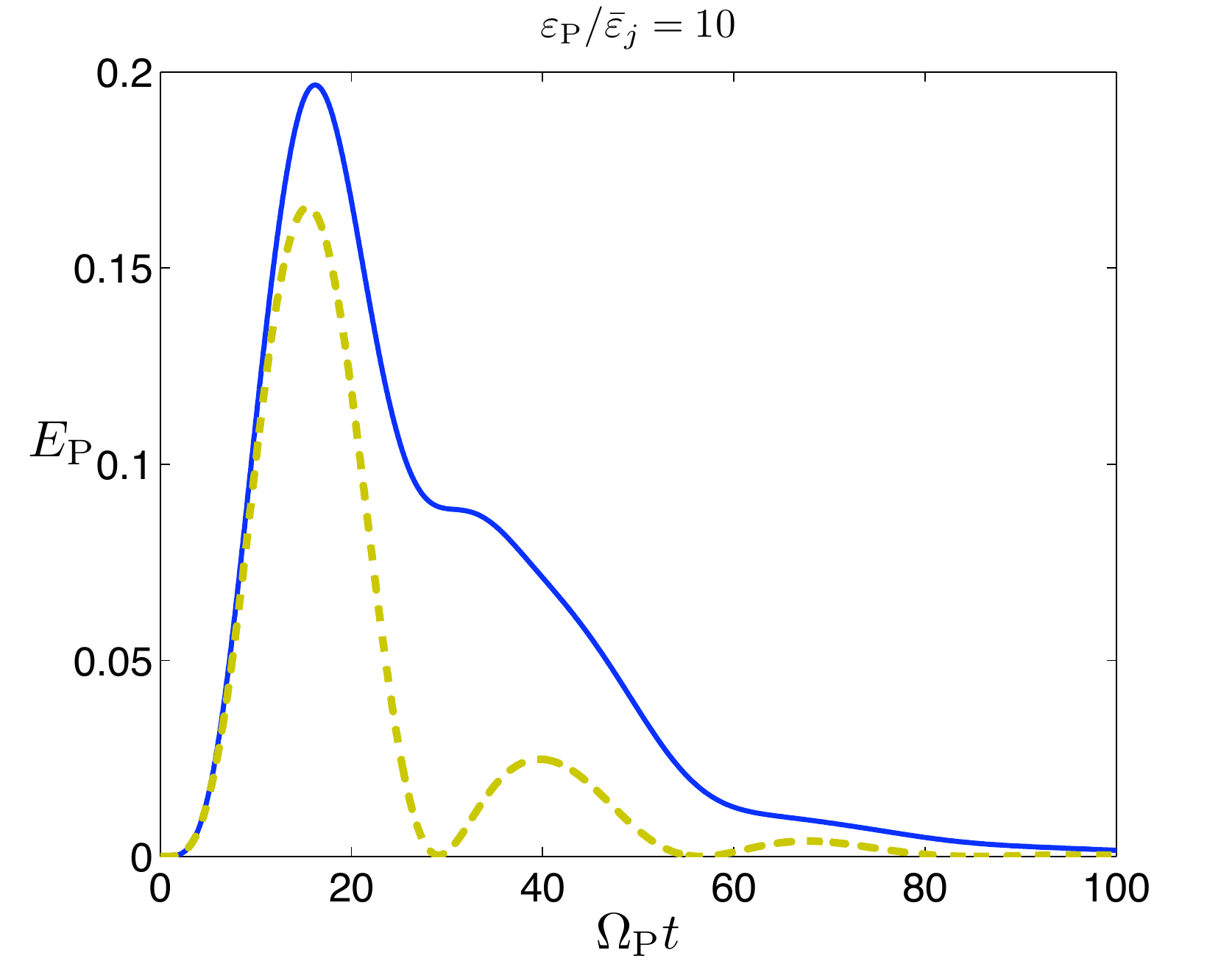}\\
			\includegraphics[width=\textwidth]{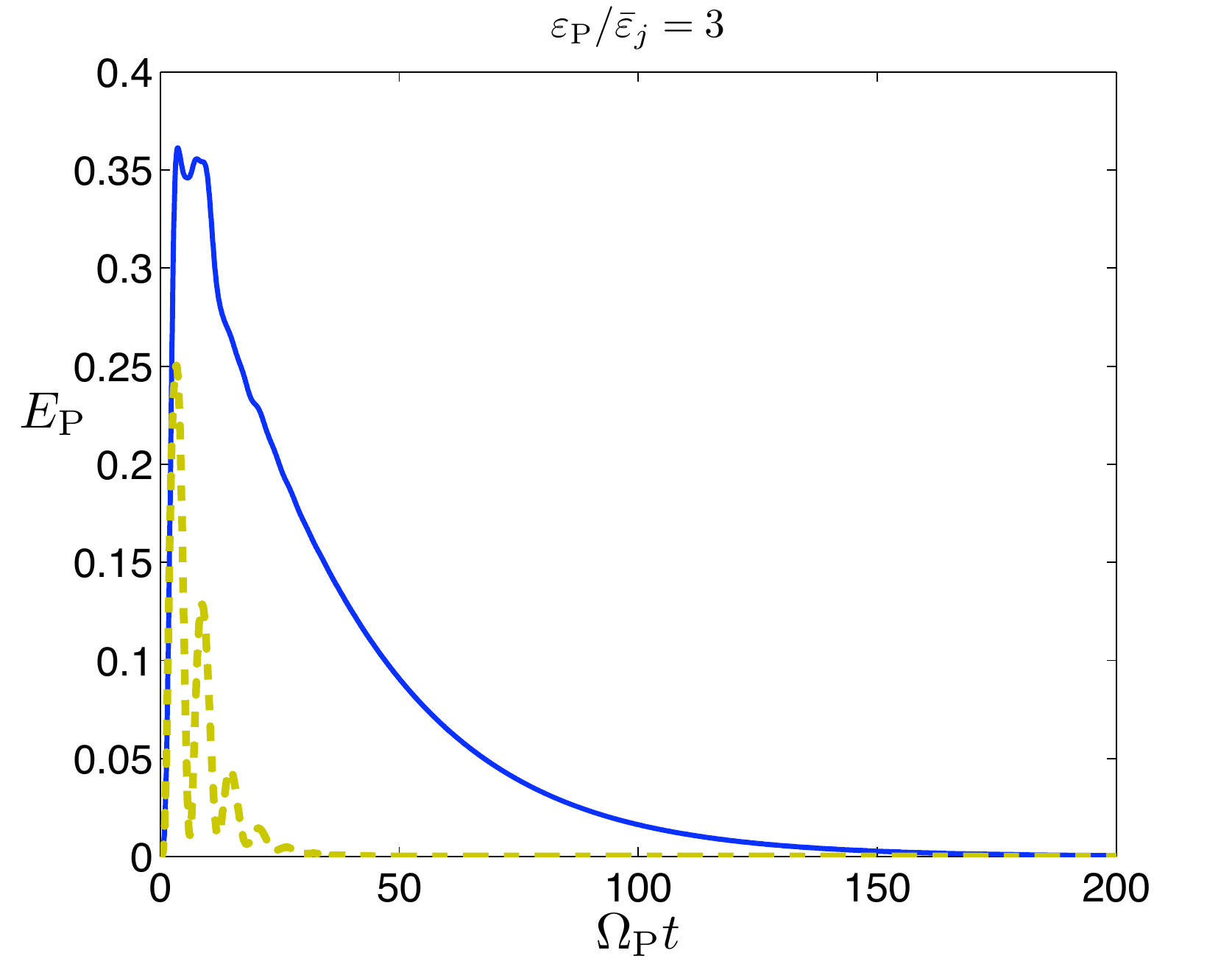}\\
			\includegraphics[width=\textwidth]{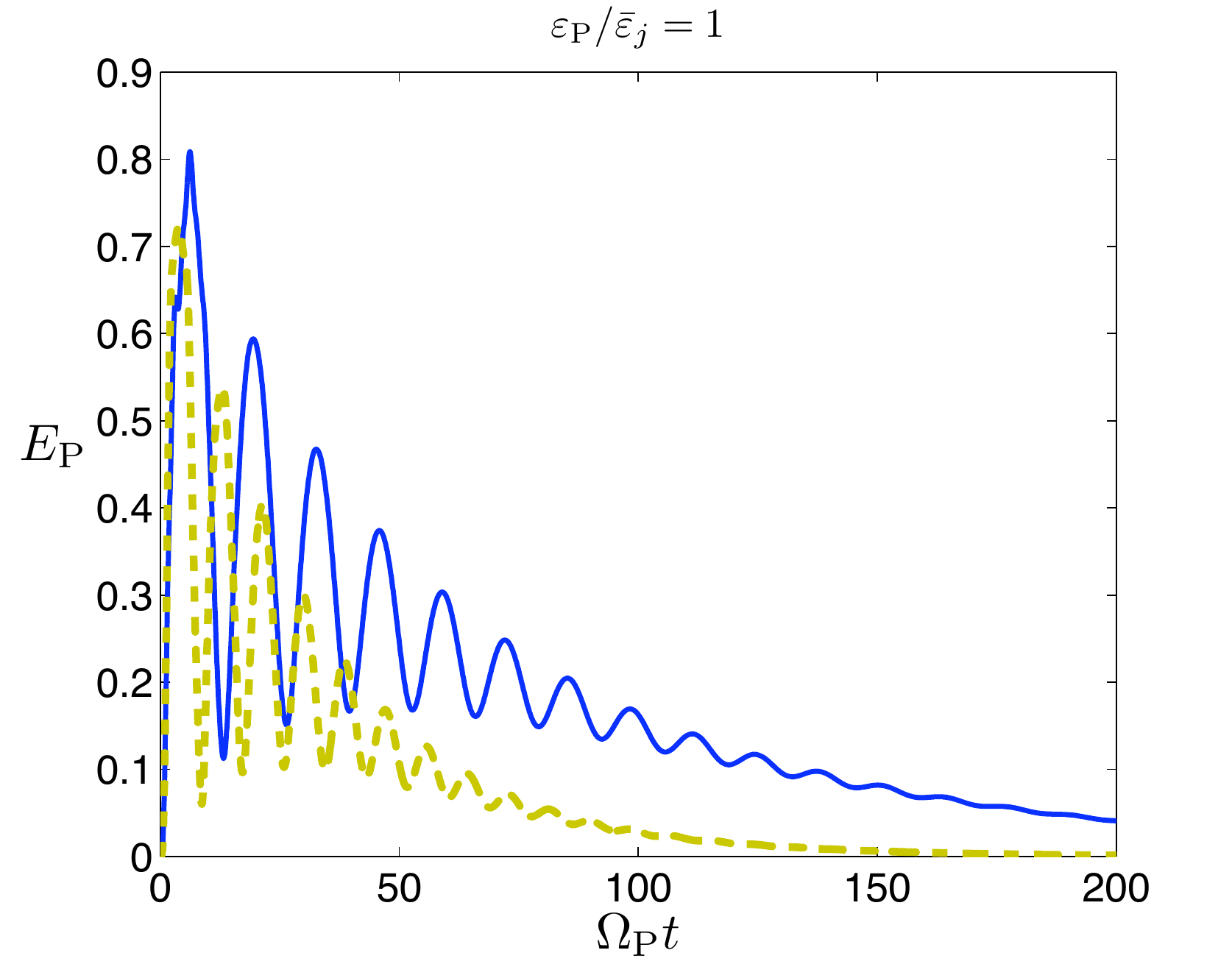}%
		\end{minipage}}
	\caption{\protect\label{fig5}
	Entanglement (logarithmic negativity) between the probe qubits
	for $\varepsilon_\mathrm{P}/\bar{\varepsilon}_j = 10$ (top row),
	$3$ (second row) and $1$ (third row). 
	The cases $\mu=0$ (solid blue line) and $\mu=\nu$ (dashed gold line) are
	qualitatively distinguishable unless both
	$\bar{\varepsilon}_j/\varepsilon_\mathrm{P} = 1$ and $\tan\bar{\theta}_j > 1$.
	Note that the upper limits of the axes change between plots.
	Same data sets as in \protect\fref{fig3}.
	}
\end{figure}

	Unlike the power spectrum (\fref{fig3}), the probe entanglement
    $E_\mathrm{P}(t)$ remains useful for distinguishing $\mu=0$ from $\mu\neq 0$
    for strong local fields in the TLFs where $\tan\bar{\theta}_j > 1$.
    Figure \ref{fig5} shows the probe entanglement as a function of time
    for the same data sets as in \fref{fig3}.
    It is clear that entanglement is generated within the probe for some time, before
    the TLF decoherence causes it to dissipate. As argued before, this loss of generated probe
    entanglement occurs faster for highly connected TLFs with $\mu/\nu=1$,
    as one might expect, even without no explicit mention of monogamy constraints, since these TLF-TLF connections provide more links
    between the probe qubits and the TLF decoherence channels. This faster dissipation of generated entanglement for
    highly-connected TLFs allows us to distinguish between $\mu/\nu=0$ and
    $\mu/\nu=1$ after 10 to 50 probe qubit cycles, depending on the TLF parameters.
    Similar conclusions can be drawn for $\tan\bar{\theta}_j = 1$.

\subsection{Discussion of the results}\label{sec:discussion}
Since we are performing local measurements on each probe qubit, one might ask
if there are any advantages of using a double-qubit probe.
An obvious advantage is that entanglement within the probe becomes
an accessible quantity that is not possible in a single-qubit probe.
This is important for detecting the TLF-TLF connectivity, as we have
seen that this task was achievable over a wider range of TLF parameters
using the probe entanglement than the power spectra of the probe
magnetization (which yielded qualitatively similar results for both
types of probes).

In section \ref{sec:connectivity} we were able to distinguish between
composite spin-boson environments with high connectivity from those with
low connectivity.  The parameter ranges for which our findings were
robust are: $\varepsilon_\mathrm{P}/\bar{\varepsilon}_j = 1,3,10$
(tunable), $1/3 \leq \tan\bar{\theta}_j \leq 3$.  Other parameters
($\Gamma$, etc) are restricted to lie within the ranges that ensure
validity of the master equation --- see \ref{sec:MEderive}.

For all numerical calculations in this paper we have assumed effectively
zero-temperature bosonic baths where
$\bar{n}_j = [\exp(\hbar\Omega_j/k_\mathrm{B}T) - 1]^{-1} \ll 1$.
The relevant frequencies for experiments with Josephson qubits are in the
vicinity of 10 GHz \cite{NakPasTsaNAT99,SimPRL04,NeeleyNATP08,LupARXIV08},
with cryostat temperatures of the order of 30 mK \cite{NakPasTsaNAT99}.
These values give $\bar{n}\sim 0.1$, so the low-temperature approximation
is good.

In general, determining the probe entanglement would require full
quantum-state tomography.  Here we consider estimating the probe
entanglement from measurements less costly than full quantum state
tomography, as described in \cite{AudPleNJP2006} (and references
within).  The result is lower bounds on the entanglement --- we refer the
reader to \cite{AudPleNJP2006} for details.

\begin{figure}
	{\centering
		\subfigure[~$\mu = 0$.]%
			{\includegraphics[width=0.45\textwidth]{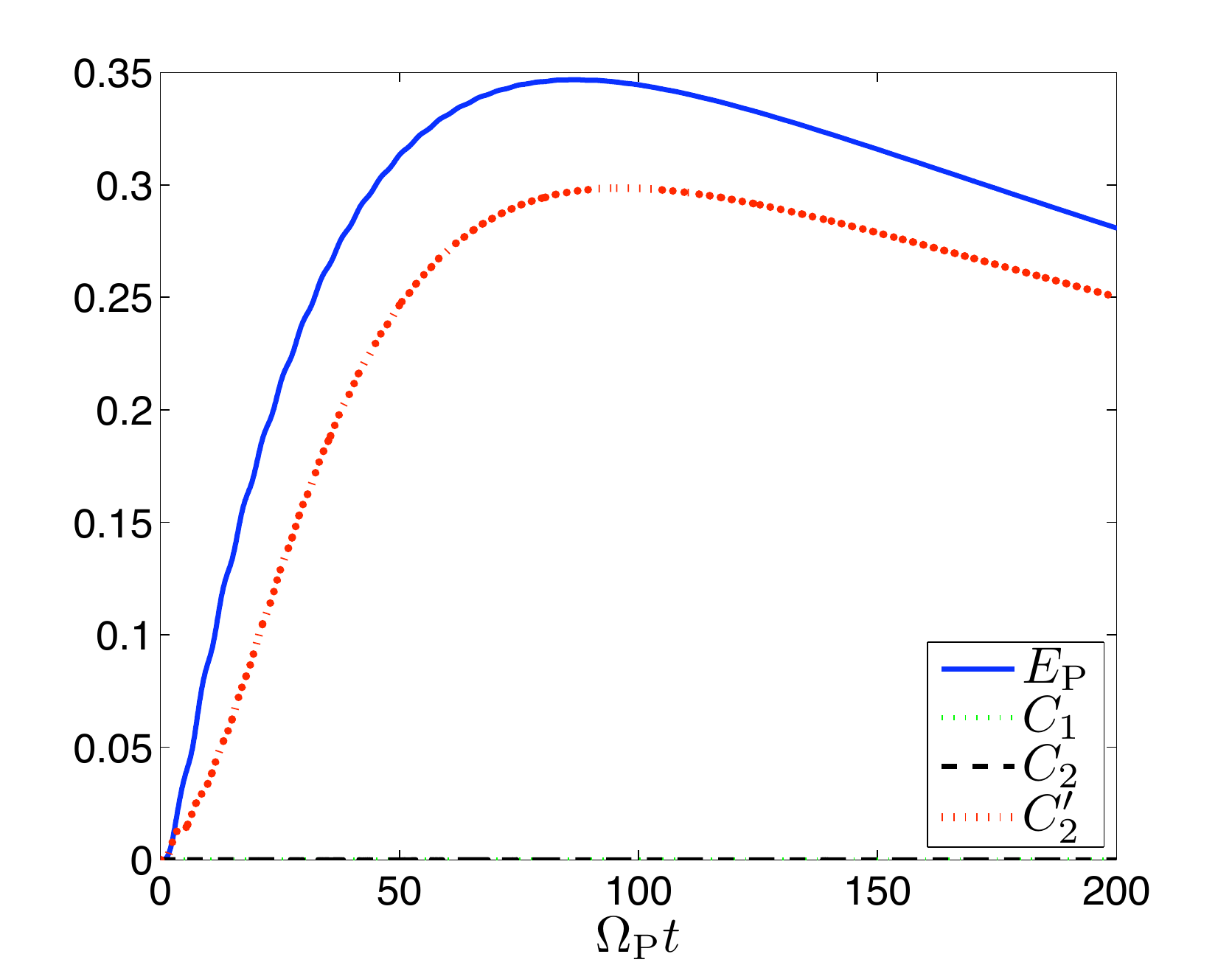}}%
		\subfigure[~$\mu = \nu$.]%
			{\includegraphics[width=0.45\textwidth]{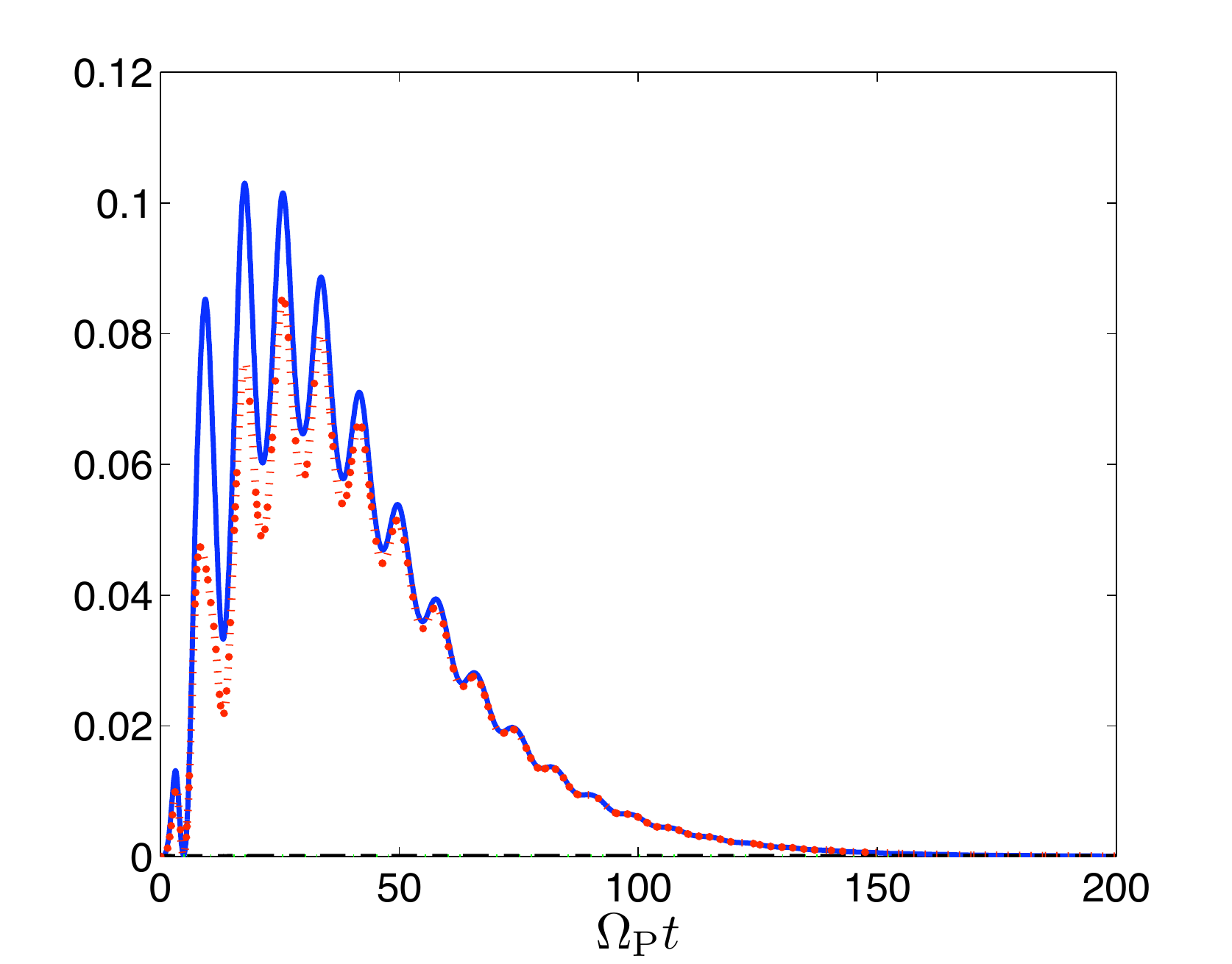}}
	\caption{\protect\label{fig6}
	Lower bounds on the probe entanglement as described in \protect\cite{AudPleNJP2006}.
	Only the optimal bound $C^\prime_2$ (dotted red line) provides a good
	approximation to the probe entanglement.
	Parameters are $\tan\bar{\theta}_j = 1/3$,
	$\varepsilon_\mathrm{P}/\bar{\varepsilon}_j=1$, and $\ket{\psi(0)}=\ket{++}$.}
	}
\end{figure}

For all parameter regimes considered in this paper we found that
one of the lower bounds given in \cite{AudPleNJP2006} --- the optimal one
given in (\ref{eq:Cprime}) --- provided a remarkably good approximation to
the probe entanglement for all times.
The other lower bounds $C_{1,2}$ in \cite{AudPleNJP2006} didn't approximate
the probe entanglement well for any time.
An example is shown in figure \ref{fig6}.  The solid line is the
entanglement within the probe (logarithmic negativity), and the dotted
line shows the lower bound given by
\begin{equation}
    C^{\prime}_2(\rho_\mathrm{P}) = \max [ 0 , \log_2(1 + |\lambda_1| + |\lambda_2| + |\lambda_3|) - 1] ,
    \label{eq:Cprime}
\end{equation}
where $\lambda_{1,2,3}$ are the eigenvalues of the matrix
\begin{equation}
    \Lambda = \left(
        \begin{array}{ccc}
            c^{xx} & c^{xy} & c^{xz} \\
            c^{yx} & c^{yy} & c^{yz} \\
            c^{zx} & c^{zy} & c^{zz}
        \end{array}
                        \right) .
    \label{eq:Lambda}
\end{equation}
The matrix $\Lambda$ is formed from probe observables
$c^{ij}=\Tr [\s_i^\mathrm{A} \otimes \s_j^\mathrm{B} \rho_\mathrm{P}]$
($i,j = x,y,z$).  Note that $c^{ij} = c^{ji}$ due to the symmetry of the problem.

\section{Decoherence of entangled states}\label{sec:entdecay}
Entanglement has been identified as a key resource for quantum information processing.
It is therefore important to study the loss of entanglement induced by coupling with an
environment, as this coupling is generally unavoidable.  In this section we reconsider
our double-qubit probe as a double-qubit \textit{register} (DQR) interacting with the same
spin-boson environment (four damped TLFs) as above.  Starting the DQR in an entangled
pure-state, and the TLFs in their ground state (as before) with weak local fields
$\tan\bar{\theta}_j < 1$, we numerically investigate the behaviour of the logarithmic
negativity as a function of time.
Specifically, we consider the lifetime of distillable entanglement (for which the
logarithmic negativity is an upper bound).
We compare our results to previous studies of entanglement decay
\cite{aolita:080501,hein:032350}, 
all of which used rather less sophisticated models for the environment.
Nevertheless, we find some qualitative similarities between our results and previous work.

\begin{figure}
	\begin{center}
		\includegraphics[width=0.45\textwidth]{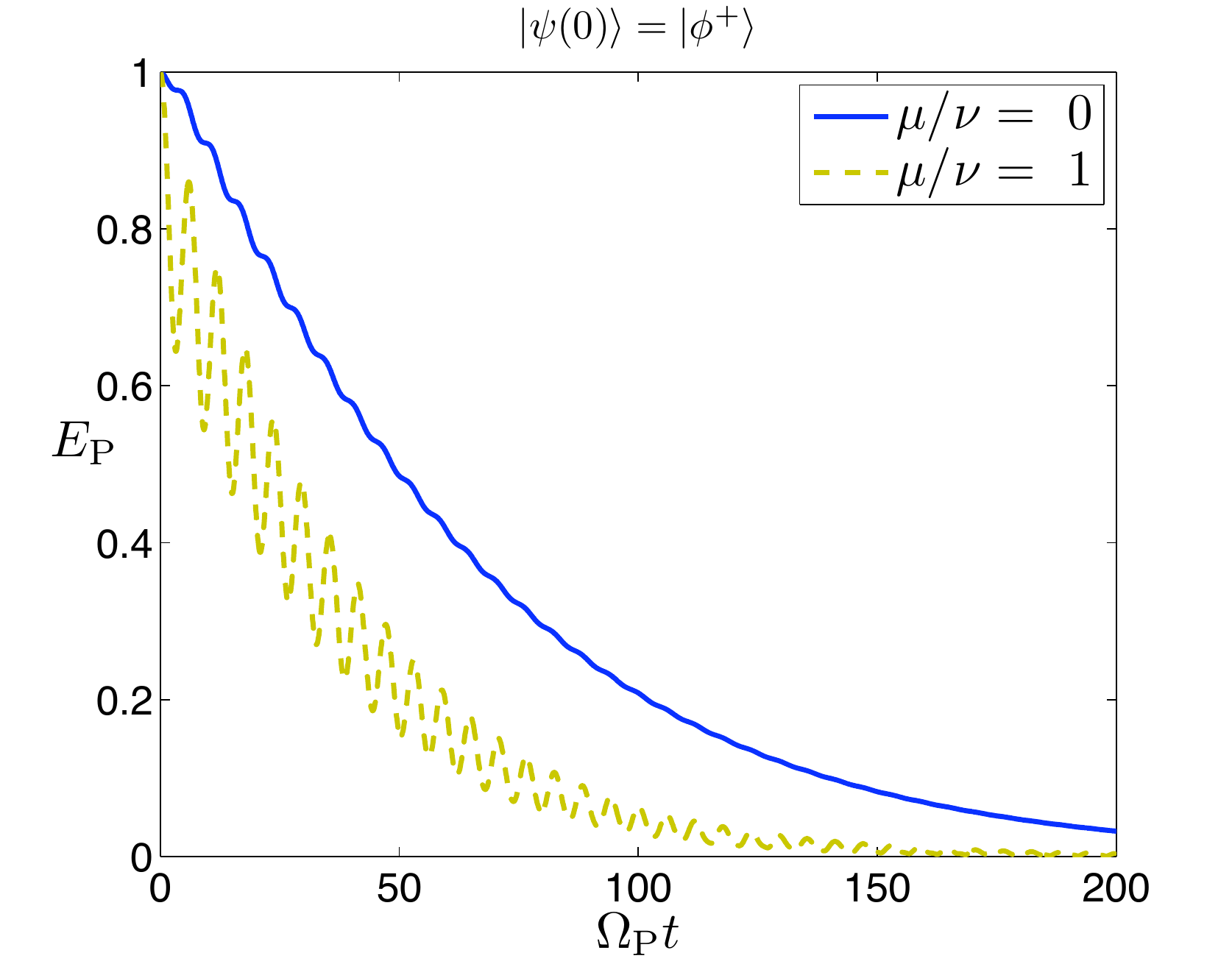}%
		\includegraphics[width=0.45\textwidth]{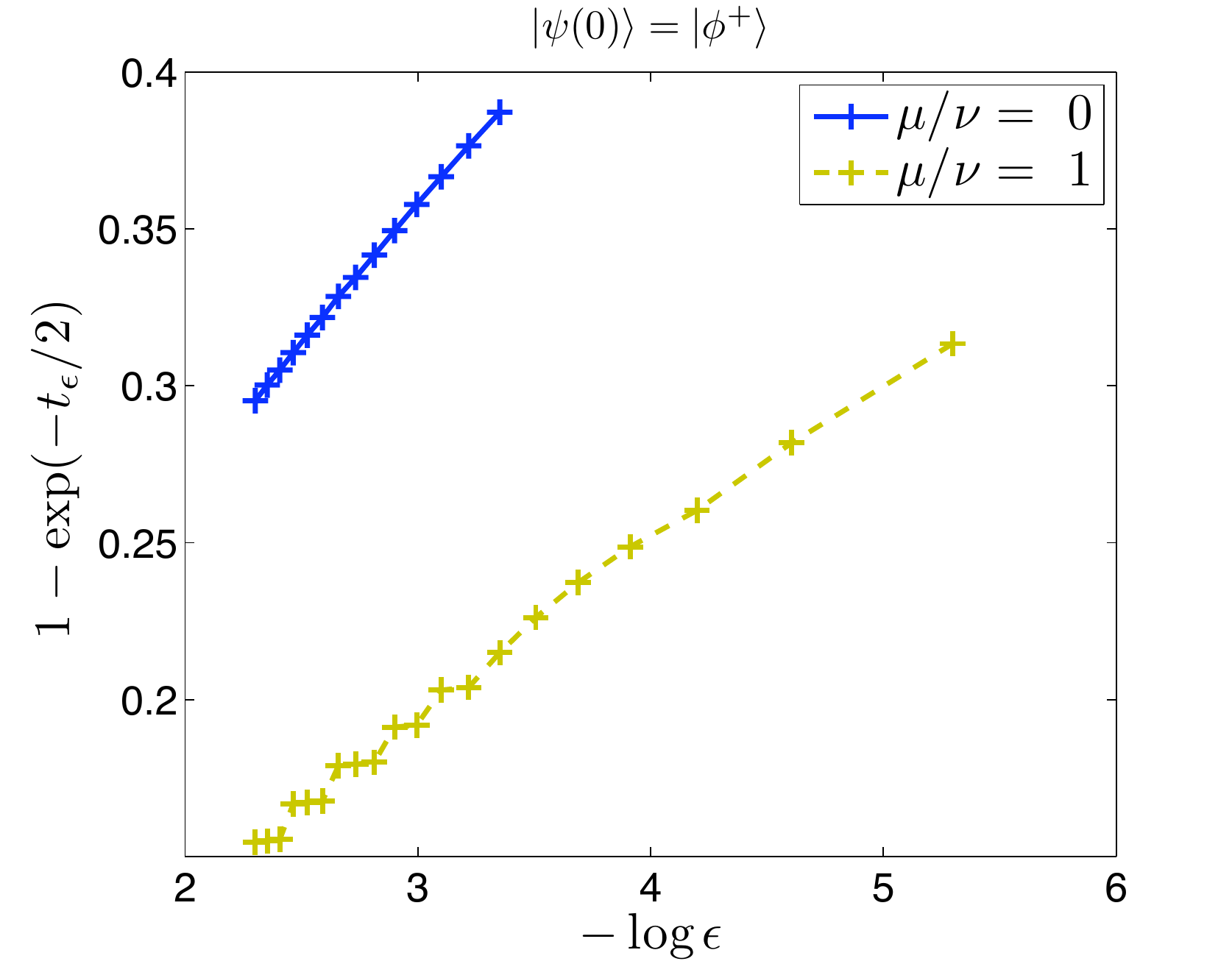}\\
		\includegraphics[width=0.45\textwidth]{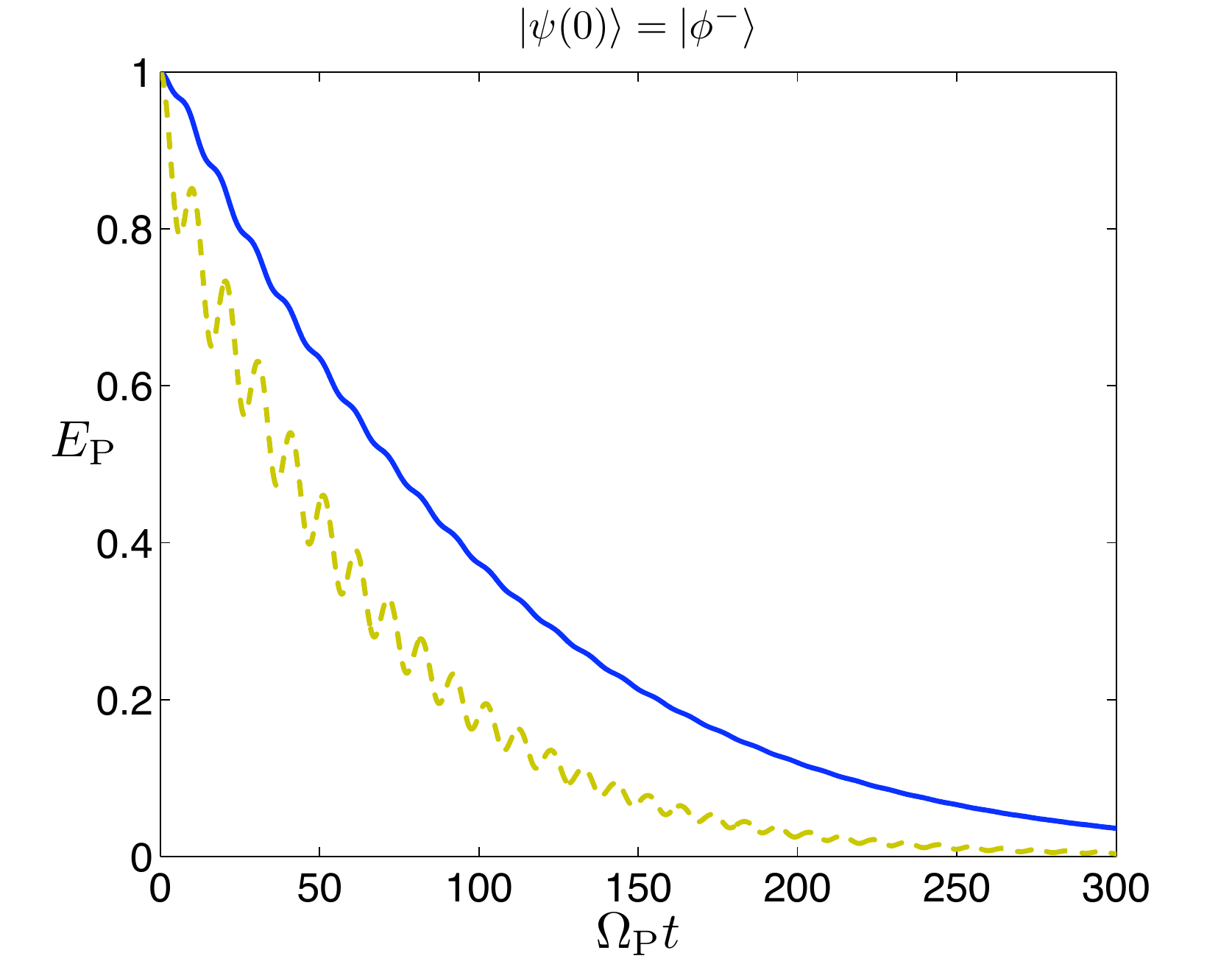}%
		\includegraphics[width=0.45\textwidth]{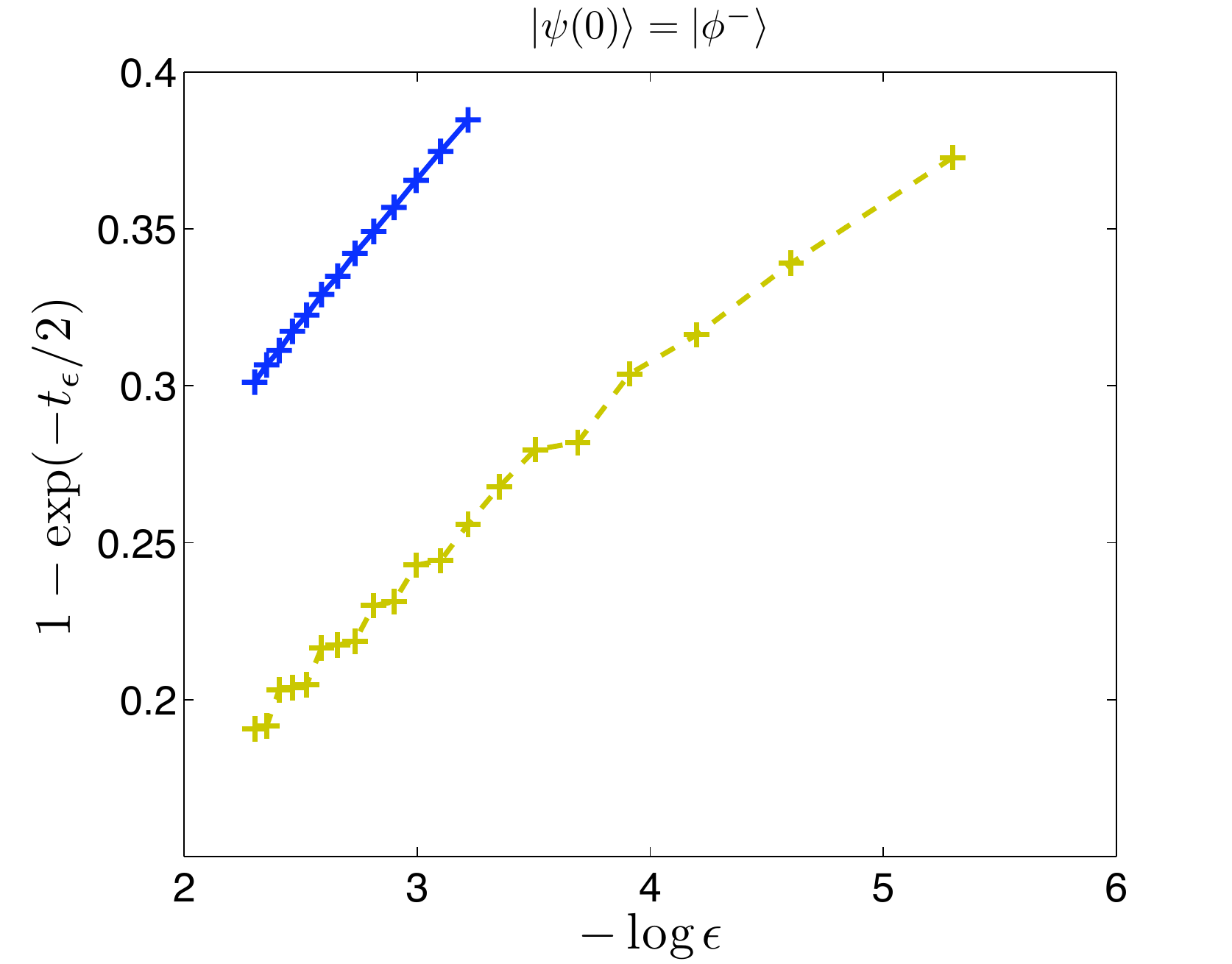}
	\end{center}
		\caption{\protect\label{fig7}
		Decay of entanglement (logarithmic negativity) in a double-qubit register initially
		in the Bell states $\ket{\phi^+}$ (top row) and $\ket{\phi^-}$ (second row)
		for uncoupled TLFs ($\mu=0$, solid blue line) and coupled TLFs
		($\mu=\nu$, dashed gold line).
		Other parameters are $\tan\bar{\theta}_j = 1/3$ and
		$\varepsilon_\mathrm{P}/\bar{\varepsilon}_j=3$.
		See text for discussion.}
\end{figure}

Reference \cite{aolita:080501} consider multiqubit states whereby each qubit is damped by
independent baths.  We refer to this as `direct' damping, by an equilibrium environment
(a reservoir).  In our non-equilibrium 
spin-boson environment,
the damping is mediated by the TLFs and we refer to this as `indirect' damping of the DQR.
Reference \cite{aolita:080501} parameterizes time via the probability for a qubit to
exchange a quantum of energy with its bath (in the absence of pure dephasing),
$p(t) = 1 - \exp[-\gamma(2\bar{n}+1)t/2] $.  Here $\gamma$ is the zero-temperature
damping rate, and $\bar{n}$ is the mean number of excitations in the bath
($\bar{n}=0$ is zero temperature).  
When the DQR logarithmic negativity falls below an arbitrarily small fraction of its initial value,
$\epsilon\ll1$, the distillable entanglement can be considered zero.
The time at which this occurs is $t_\epsilon$.
For generalized GHZ states (requiring $>3$ qubits), and for three different types of
direct damping, \cite{aolita:080501} found that $p(t_\varepsilon) \propto -\log \epsilon$
(although they were looking at it from a slightly different perspective, as we comment
later).
Remarkably, we find numerical evidence of the same qualitative behaviour for the decay
of the $\ket{\phi^\pm}$ Bell states in a DQR (see figure \ref{fig7}).

For low temperature $T\sim$mK (as we have considered throughout this paper
unless otherwise noted), weak TLF local field $\tan\bar{\theta}_j=1/3$, and
TLFs with relatively small charge: $\varepsilon_\mathrm{P} = 3\bar{\varepsilon}_j$,
we consider the DQR to be initially in each of the four Bell states in turn:
$\ket{\phi^\pm} = (\ket{00} \pm \ket{11})/\sqrt{2}$,
$\ket{\psi^\pm} = (\ket{01} \pm \ket{10})/\sqrt{2}$.
Figure \ref{fig7} shows the DQR logarithmic negativity as a function of time, as well as
$p(t_\epsilon) = 1-\exp(-t_\epsilon/2)$ for a DQR initially in the states $\ket{\phi^\pm}$.
It is evident that $p(t_\epsilon) \propto -\log\epsilon$ for $\ket{\phi^\pm}$.
Further, we can see that interacting TLFs (the gold traces) tend to reduce the
DQR entanglement faster.  This is expected since interacting TLFs provide more
connections between the DQR and the baths.
For the chosen set of parameters (particularly $\Delta=0$ for the DQR), the
states $\ket{\psi^\pm}$ commute with $\hat{H}_\mathrm{P} + \hat{V}_\mathrm{P}$ and
so don't evolve, nor couple to the TLFs.
We found the same qualitative behaviour for
$\varepsilon_\mathrm{P}/\bar{\varepsilon}_j = 1$ and $10$ when the TLF local fields were
not strong: $\tan\bar{\theta}_j \lesssim 1$.  For strong local fields $\tan\bar{\theta}_j \gtrsim 3$,
the linear relationship $p(t_\epsilon) \propto -\log\epsilon$ did not hold in general.
This was because the probe entanglement tended to exhibit quite erratic behaviour,
such as multiple collapses and revivals.  

A comment on the previous work in \cite{aolita:080501,hein:032350} is appropriate here.
In those works, the decay of $N$-particle entanglement was considered as a function of
$N$.  In \cite{aolita:080501} it was found that $p(t)\propto -(1/N)\log\epsilon$.
Here we have fixed $N=2$ and found that $p(t)\propto -\log\epsilon$.  Our focus is
slightly different, but it is interesting that the loss of distillable entanglement
(logarithmic negativity) is qualitatively the same for direct and indirect damping
of bipartite qubit states (within the parameter regimes discussed in the previous
paragraph). It is important to remark that the coincidence with the predictions for the decoherence of multipartite states subject to independent reservoirs should not be considered as a general result given that we analyzed a very special case, which is the one of two entangled qubits in selected parameter regimes. What is relevant for our purposes is the fact that the agreement with the analytical prediction in \cite{aolita:080501} for direct decoherence points out a sharp asymmetry in the processes of entanglement "destruction" and (remote) entanglement generation in a composite environment, so that there are circumstances where the TLF systems may be essentially invisible when analyzing the decoherence of initially entangled probe states, while the presence of the TLFs would be revealed when monitoring entanglement creation in the probe.

\section{Effect of spin-boson environment on entangling gate operations}\label{sec:gates}
\begin{figure}[ht]
	\centering
	\subfigure[~$\tan\bar{\theta}_j = 1/3$.]%
		{\label{fig8a}%
		\begin{minipage}{0.45\textwidth}
			\includegraphics[width=\textwidth]{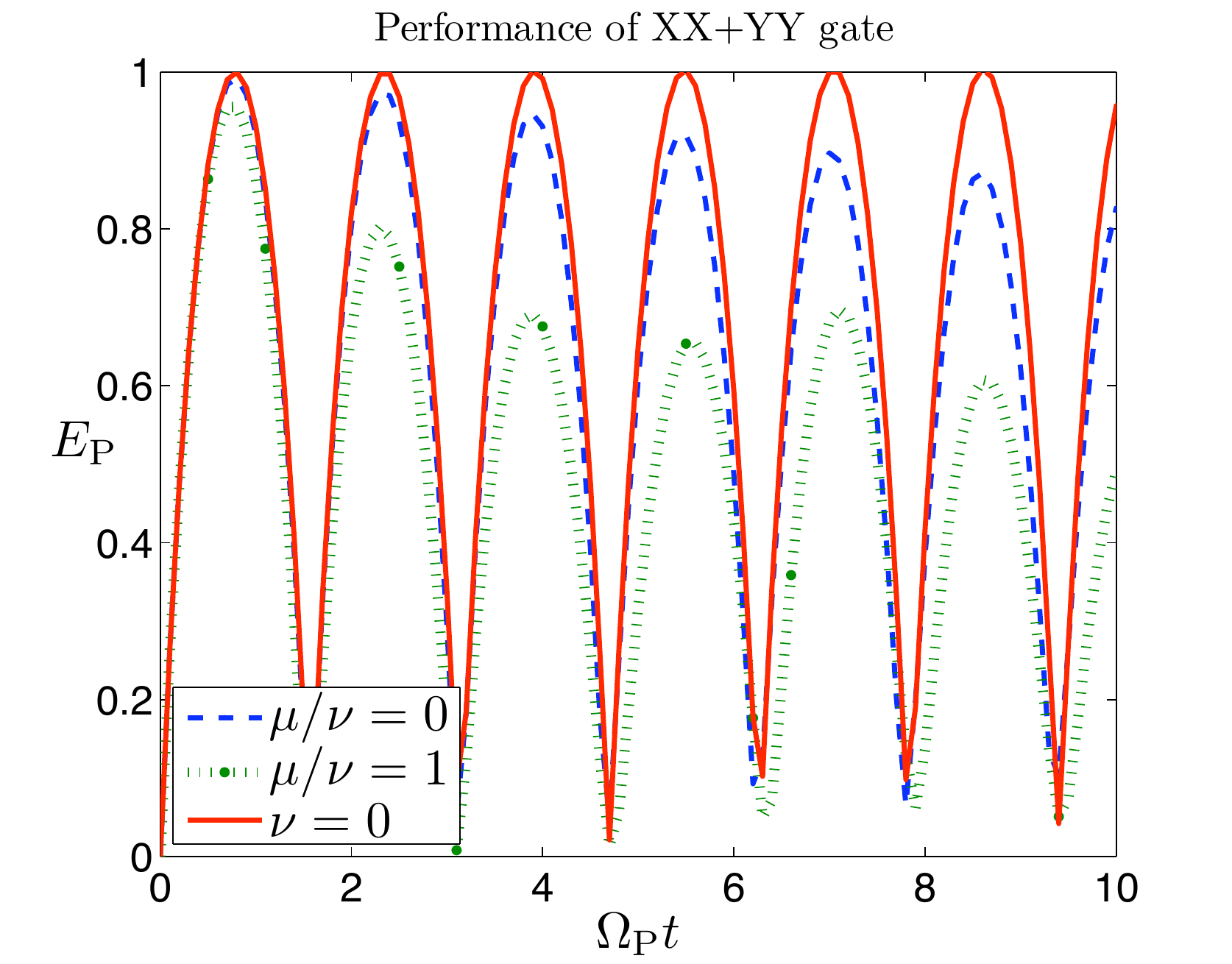}\\
			\includegraphics[width=\textwidth]{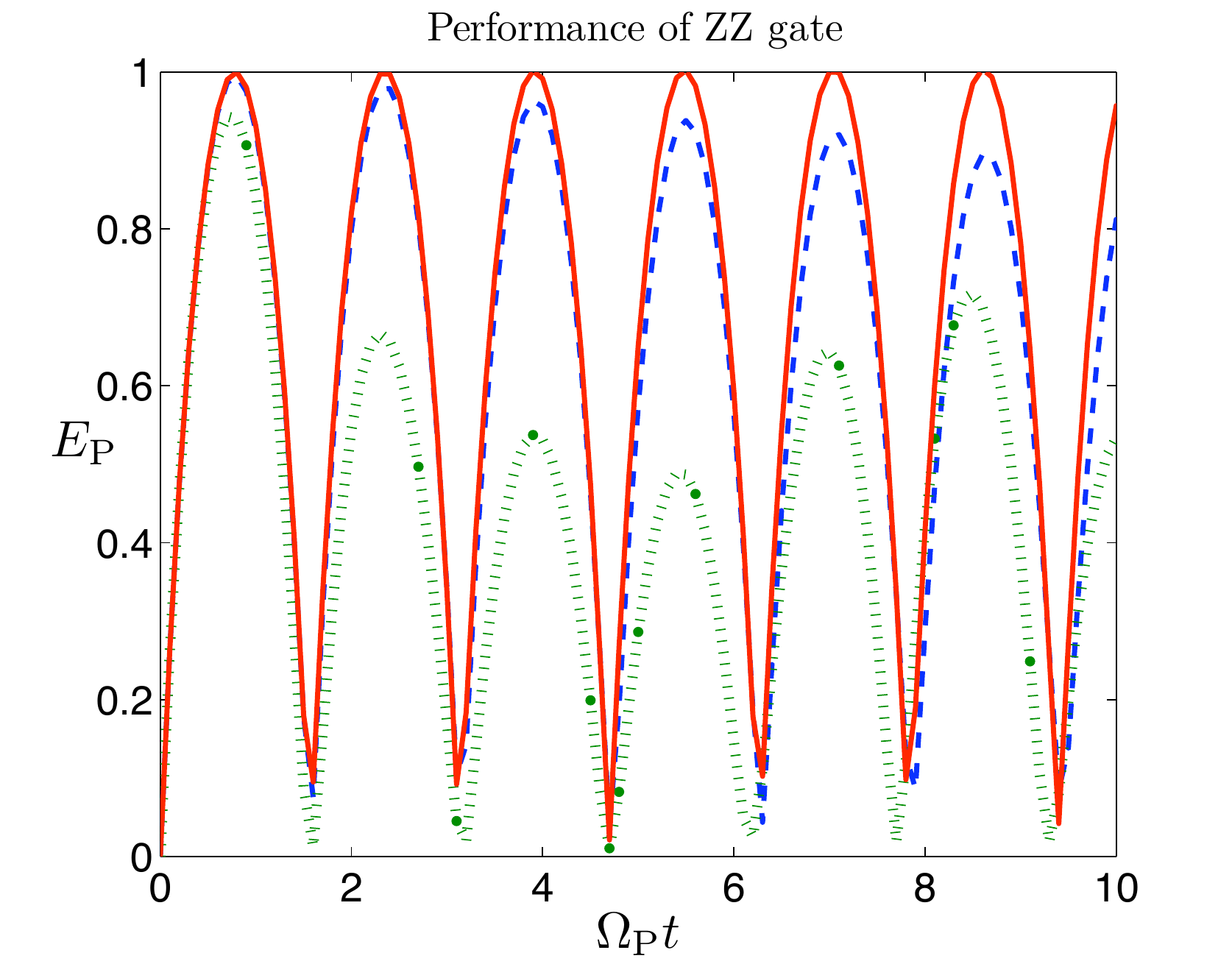}%
		\end{minipage}}%
	\subfigure[~$\tan\bar{\theta}_j = 3$.]%
		{\label{fig8b}%
		\begin{minipage}{0.45\textwidth}
			\includegraphics[width=\textwidth]{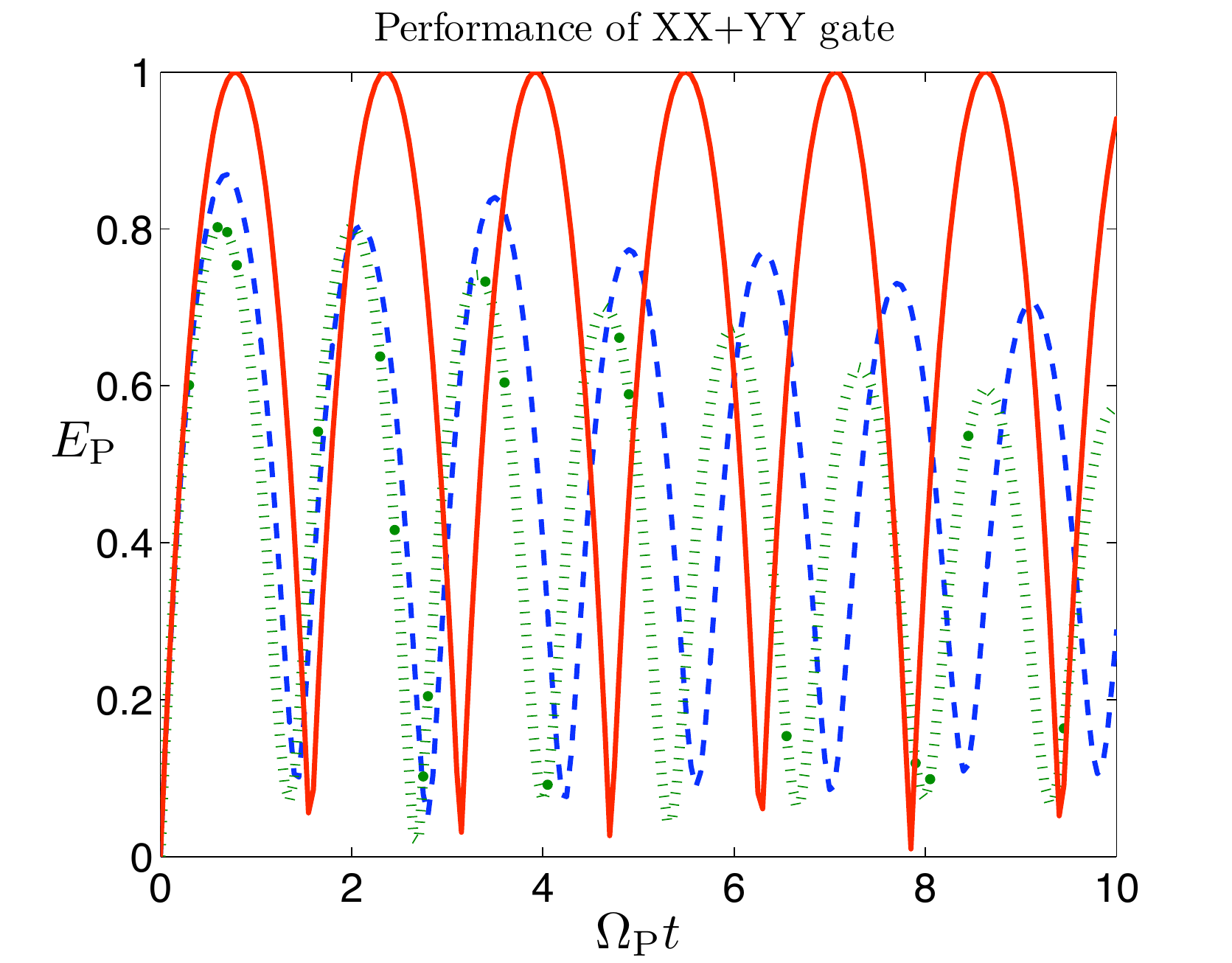}\\
			\includegraphics[width=\textwidth]{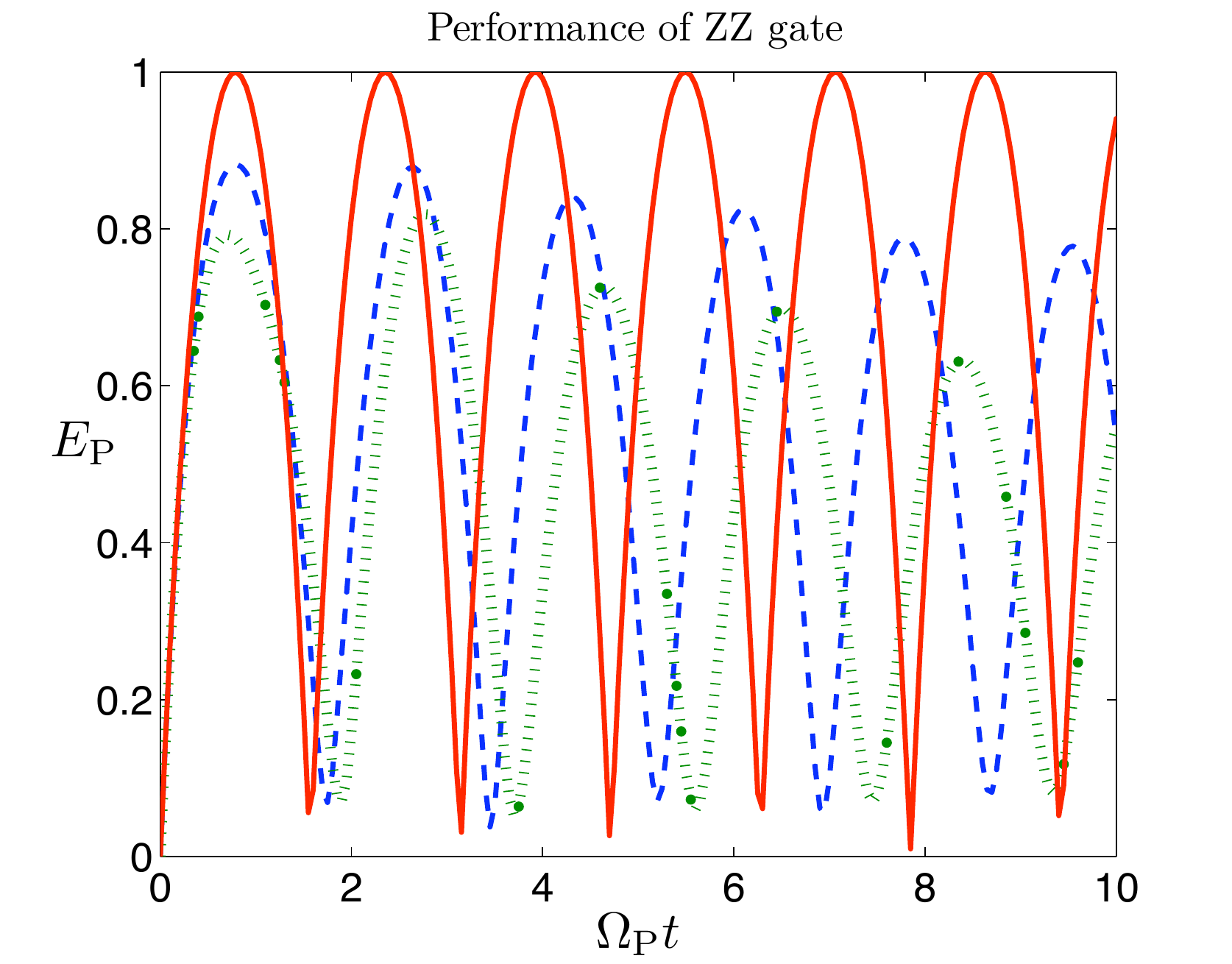}%
		\end{minipage}}%
	\caption{\protect\label{fig8}
	Entanglement (logarithmic negativity) between the register
	qubits for $\ket{\psi(0)}=\ket{++}$, $\varepsilon_\mathrm{P}/\bar{\varepsilon}_j = 1$
	generated by XX+YY (top row), and ZZ (bottom row) gates.
	The presence of TLFs (blue dashed and green dotted lines)	diminishes
	the performance of the entangling gates, as would be expected. }
\end{figure}

\begin{figure}[ht]
	\centering
	\includegraphics[width=0.65\textwidth]{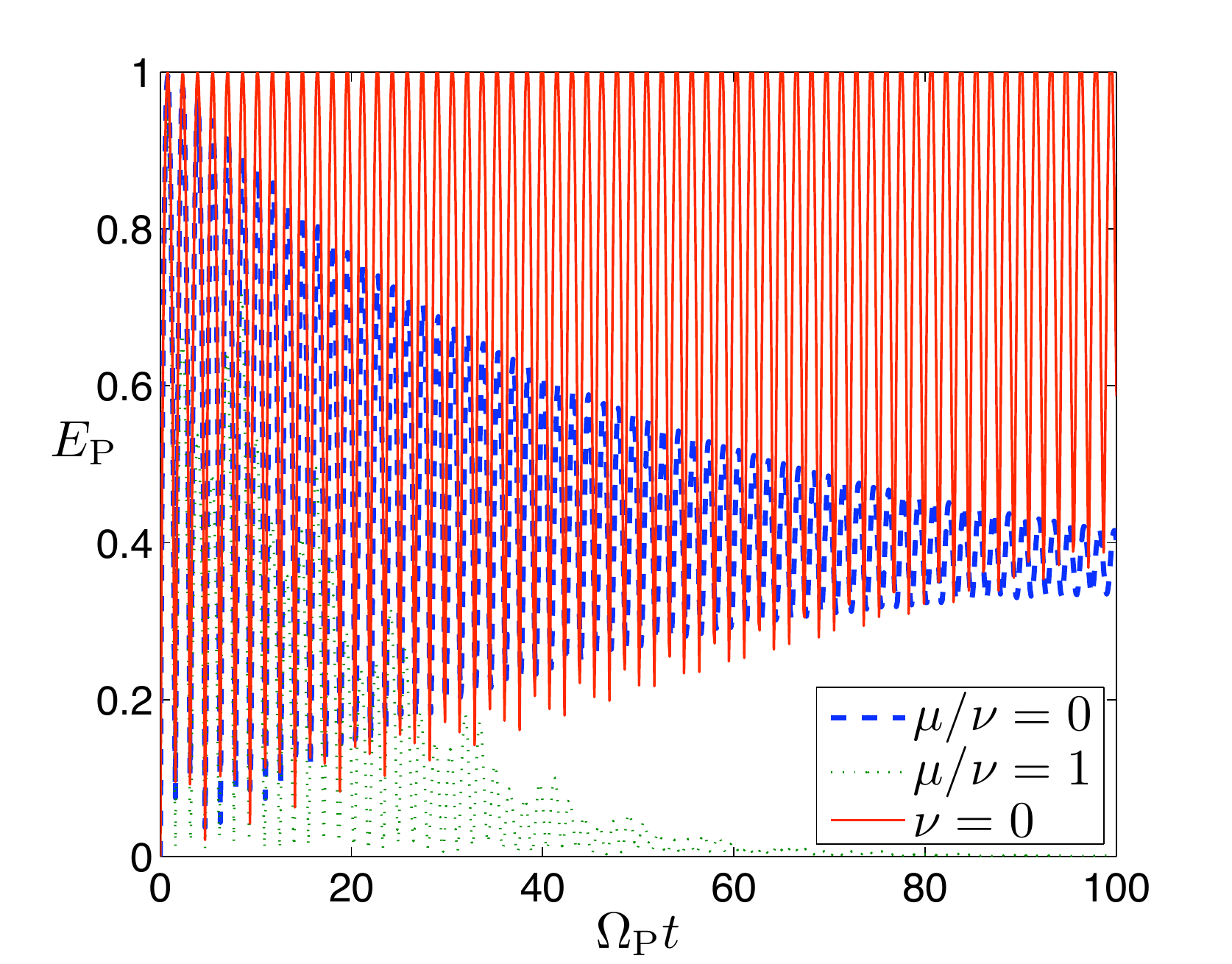}
	\caption{\protect\label{fig9}
	Long-time entanglement (logarithmic negativity) between the register
	qubits generated by a ZZ gate, for $\ket{\psi(0)}=\ket{++}$,
	$\varepsilon_\mathrm{P}/\bar{\varepsilon}_j = 1$ and $\tan\bar{\theta_j}=1/3$.}
\end{figure}

In this section we investigate the effects of the composite spin-boson environment
on the performance of entangling gates.
Starting the DQR in the separable state $\ket{\psi_\mathrm{P}(0)} = \ket{++}$,
we consider two entangling gates:
a ZZ gate ($\sz^\mathrm{A}\otimes\sz^\mathrm{B}$);
and
an XX+YY gate
[$\sx^\mathrm{A}\otimes\sx^\mathrm{B} + \sy^\mathrm{A}\otimes\sy^\mathrm{B}$].
In the ideal case there are no TLFs and bipartite entanglement (quantified again
by the logarithmic negativity) is generated between the isolated register qubits
in an oscillatory fashion as shown by the solid red curves in \fref{fig8}.
In the presence of four TLFs, the entangling gate performance is clearly reduced,
and further modified depending on the strength of the local fields in the TLFs.
This may be understood as follows. We have argued that the presence of a coherent coupling between the TLFs leads to a decrease in the
effective interaction strength between the qubits in the probe, i.e., the ``effectiveness''
of the indirect link between probe qubits is diminished, a result that can be interpreted in terms of monogamy constraints
leading to a slow down in the process of remote entanglement creation.
In the case of the entangling gates, one could perhaps argue similarly, considering
now a bipartition separating the probe qubits. The higher the connectivity in the environment,
the slower the entangling gate can operate, as illustrated in \fref{fig9}.

At longer times (the order of 100 register-qubit cycles and greater),
the DQR appears to approach an entangled steady state
for weak local fields $\tan\bar{\theta}_j = 1/3$.  This is shown in
\fref{fig9} for the ZZ gate (the same qualitative behaviour occurred
for the XX+YY gate).
We were unable to obtain analytical results to verify the presence of
steady-state entanglement in the DQR generated by either gate (ZZ or XX+YY),
when coupled to four TLFs.  Our numerical study found that entanglement
generated in the DQR was dissipated more rapidly (in less probe qubit cycles)
for strong local fields in the TLFs.

\section{Conclusion}\label{sec:conclusion}
We have considered superconducting qubits which are subject to decoherence
dominated by low-frequency noise thought to be produced by interactions
with a small number of defects/impurities.
We model these impurities as coherent two-level fluctuators (TLFs) that
are under-damped by baths of bosonic modes (e.g. phonons).
We probed such a composite spin-boson environment by making measurements
on a pair of noninteracting qubits that each interact directly with the
TLFs.
Our extensive numerical study revealed that the presence or absence of
coherent coupling (\textit{connectivity}) between the TLFs can be
discriminated in two ways:
from the estimated power spectrum of the probe magnetization (requiring
relatively long-time measurements)
and from entanglement generated within the probe, mediated by the spin-boson
environment. We argue that entanglement monogamy considerations
\cite{CofKunWooPRA00} allow interpretation of our results in terms of an effective decrease in the remote interaction
strength when the environment is connected, which yields to a creation
of quantum correlations on a much larger time scale as compared with the uncoupled fluctuator case.

We also showed that this remotely generated entanglement can be well-estimated
by a lower bound \cite{AudPleNJP2006} that requires less experimental
effort than the full quantum state tomography required to evaluate the entanglement.
The upshot is that this connectivity of the TLFs should be discernible
using tractable measurements in a real experiment.
This result is important for studies of quantum-mechanical phenomena
in Josephson devices (including quantum computing) where it is desirable
to minimize the effects of decoherence, for which the TLF-TLF connectivity
may play a significant role \cite{YuanPRB08}.

When considering the effects of the spin-boson environment on a
double-qubit register initially prepared in a maximally-entangled (Bell)
state, we showed that the presence of TLFs may be unnoticeable in certain parameter regimes, in the sense that
entanglement degradation there is well approximated by the same decrease law as for direct decoherence. This fact emphasizes the
possible usefulness of monitoring the reverse process of entanglement generation for environmental probing.

The presence of TLF-TLF coupling also
showed a reduction in the performance of entangling gate operations performed on the
register. Our extensive numerical study
can be supplemented by an analytical treatment of a simpler situation
valid for short times, where TLF decoherence can be ignored.
These results will be presented elsewhere \cite{UHinprep}.

\ack
We are very grateful to Simone Montangero for discussions on several aspects of the simulations and to Shash Virmani for
discussions on monogamy constraints.
This work was supported by the EU through
the STREP project CORNER,
the Integrated project on \emph{Qubit Applications} QAP, and
the Integrated project EuroSQIP.
AR acknowledges support from a University of Hertfordshire Fellowship.

\appendix
\section{Derivation of the master equation}\label{sec:MEderive}
Consider a single charge qubit ``system'' coupled to a finite number of
independent charged impurities that fluctuate coherently between two
configurations.  These coherent two-level fluctuators (TLFs) are coupled to
independent bosonic baths (phonons in the substrate, for example) that produce
damping.  The total Hamiltonian is the sum of free Hamiltonians for the single
qubit, the TLFs, and the baths, and the interaction Hamiltonians:
\[
\hat{H} = \hat{H}_{\mathrm{sq}} + \hat{H}_{\mathrm{TLF}} + \hat{H}_{\mathrm{B}}+
\hat{V}_{\mathrm{sq-TLF}} + \hat{V}_{\mathrm{TLF-B}} .
\]
The free Hamiltonians are
$\hat{H}_{\mathrm{sq}} = (\varepsilon\sigz + \Delta\sigx)/2$,
$\hat{H}_{\mathrm{TLF}} = \sum_j(\varepsilon_j\sigz^{(j)} + \Delta_j\sigx^{(j)})/2$,
$\hat{H}_{\mathrm{B}} = \sum_{j,\ell}\omega_\ell \hat{a}^\dagger_{\ell,j} \hat{a}^{\phantom{\dagger}}_{\ell,j}$ (the baths
can be assumed to be identical so $\omega_{\ell,j}=\omega_\ell$).
The coupling Hamiltonians are
$\hat{V}_{\mathrm{sq-TLF}} = \sum_j \nu_j \sigz^{(j)}\sigz$ (Coulomb interactions),
$\hat{V}_{\mathrm{TLF-B}} = \sum_{j,\ell} \lambda_\ell \sigz^{(j)}
( \hat{a}^{\phantom{\dagger}}_{\ell,j} + \hat{a}^\dagger_{\ell,j})$
(similarly to $\omega_\ell$, the couplings are assumed to be independent of the baths:
$\lambda_{\ell,j} = \lambda_\ell$).
We denote the charge-basis Pauli operators by $\sig_\mathrm{x,y,z}$.  At this stage the
TLFs are not interacting.

The TLF-related energies $\varepsilon_j$, $\Delta_j$ and $\nu_j$ are randomly distributed
following independent distributions discussed in \cite{ShnetalPRL05}.  The details of
these distributions are critical for realizing the experimentally observed $1/f$ noise
spectrum of the qubit voltage/bias.

We refer to the eigenbases of $\hat{H}_{\mathrm{sq}}$ and $\hat{H}_{\mathrm{TLF}}$ as
the pseudo-spin bases, and denote the corresponding Pauli operators as $\s_\mathrm{x,y,z}$.
The Hamiltonians in the pseudo-spin basis are
\begin{eqnarray*}
\hat{H}_{\mathrm{sq}} &=& \frac{1}{2}\Omega{}\sz\\
\hat{H}_{\mathrm{TLF}} &=& \frac{1}{2}\sum_j\Omega_j\sz^{(j)}\\
\hat{V}_{\mathrm{sq-TLF}} &=& \sum_j\nu_j(\cos\theta_j\sz^{(j)} - \sin\theta_j\sx^{(j)}) (\cos\theta\sz - \sin\theta\sx) \\
\hat{V}_{\mathrm{TLF-B}} &=& \sum_{\ell,j}\lambda_{\ell}(\cos\theta_j\sz^{(j)}-\sin\theta_j\sx^{(j)})\left(a^{\phantom{\dagger}}_{\ell,j}+a^\dagger_{\ell,j}\right)
\end{eqnarray*}
where $\tan\theta_j \equiv \Delta_j/\varepsilon_j$.
In order to obtain the Master equation we move to an interaction picture with respect
to the Hamiltonian
\[
\hat{H}_0=\hat{H}_{\mathrm{sq}}+\hat{H}_{\mathrm{TLF}}+\hat{H}_{\mathrm{B}}+\hat{V}_{\mathrm{sq-TLF}} .
\]
In this picture the evolution equation for the total system is:
\[
\frac{d\tilde{\rho}_{\mathrm{tot}}(t)}{dt} = -\rmi [\tilde{V}_{\mathrm{TLF-B}}(t),\tilde{\rho}_{\mathrm{tot}}(t)]
\]
where
\begin{eqnarray*}
\tilde{\rho}_{\mathrm{tot}}(t)&=&\exp(i\hat{H}_0t)\rho_{\mathrm{tot}}\exp(-\rmi\hat{H}_0t) ,\\
\tilde{V}_{\mathrm{TLF-B}}(t)&=&\exp(i\hat{H}_0t)\hat{V}_{\mathrm{TLF-B}}\exp(-\rmi\hat{H}_0t) .
\end{eqnarray*}
By iterating once the above equation as usual \cite{OpenQS,CarmichaelSMQO1} we obtain
\[
\frac{d\tilde{\rho}(t)}{dt}=-\Tr_{\mathrm{env}}\int_0^tdt'[\tilde{V}_{\mathrm{TLF-B}}(t),[\tilde{V}_{\mathrm{TLF-B}}(t'),\tilde{\rho}_{\mathrm{tot}}(t')]]
\]
where $\Tr_{\mathrm{env}}$ denote the trace over all of the baths.
By assuming a factorized initial state of the form
$\rho{_\mathrm{tot}}(0)=\rho(0)\bigotimes_j\rho_{b,j}$, with $\rho(0)$ the initial
state of the ``qubit + TLFs'' and $\rho_{b,j}$ a thermal state of the $j$th bath,
we may make a Born approximation in the coupling constants $\lambda_{\ell}$.
The evolution equation becomes
\[
\frac{d\tilde{\rho}(t)}{dt}\simeq-\Tr_{\mathrm{env}}\int_0^tdt'[\tilde{V}_{\mathrm{TLF-B}}(t),[\tilde{V}_{\mathrm{TLF-B}}(t'),\tilde{\rho}(t')\bigotimes_j\rho_{b,j}]] .
\]
Since the baths are independent and they are all in a thermal state (diagonal in the
number basis), by inserting the expression for
$\tilde{V}_{\mathrm{TLF-B}}(t)$ it is easy to check that
\begin{eqnarray*}
\frac{d\tilde{\rho}(t)}{dt}= -\Tr_{\mathrm{env}}\sum_{\ell,j}\int_0^tdt' \Bigg[ \lambda_{\ell}(\cos\theta_j\tilde{s}^{(j)}_\mathrm{z}(t)-\sin\theta_j\tilde{s}^{(j)}_\mathrm{x}(t))
(\tilde{a}_{\ell,j} + \tilde{a}^\dagger_{\ell,j})\Bigg.,\nonumber \\
\Bigg.\bigg[\lambda_{\ell}(\cos\theta_j\tilde{s}^{(j)}_\mathrm{z}(t')-\sin\theta_j\tilde{s}^{(j)}_\mathrm{x}(t'))
(\tilde{a}_{\ell,j} + \tilde{a}^\dagger_{\ell,j})~,~\tilde{\rho}(t')\bigotimes_{j}\rho_{b,j}\bigg]\Bigg] .
\end{eqnarray*}
This is basically a sum of the expressions obtained in the standard derivation \cite{OpenQS,CarmichaelSMQO1}
for the one qubit case.  Now we make the next crucial assumption in the derivation.
Assuming weak qubit-fluctuator coupling whereby $\nu_j\ll\Omega_j$
(as in the experiments of \cite{NakPasTsaNAT99,NeeleyNATP08,LupARXIV08} --- see below),
we may approximate the Heisenberg operators of the TLFs in the interaction picture as:
\begin{equation}\label{ApproxExp}
\tilde{s}(t)\simeq\exp(i\hat{H}_{00}t)\s\exp(-\rmi\hat{H}_{00}t) ,
\end{equation}
where $\hat{H}_{00}=\hat{H}_{\mathrm{sq}}+\hat{H}_{\mathrm{TLF}}$ 
is the sum of the free Hamiltonians.  For example, we find that
$\tilde{s}_\pm^{(j)}(t) = \s_\pm^{(j)} e^{\pm i \Omega_jt}$.

One may question the validity of our assumption of weak qubit-fluctuator coupling.
For guidance we consider the experiments of
\cite{NakPasTsaNAT99} (charge qubit),
\cite{NeeleyNATP08} (phase qubit)
and \cite{LupARXIV08} (flux qubit),
where $\nu_j \lesssim 0.1 GHz$ and $\Omega_j \sim \Omega \sim GHz$,
which falls within this weak-coupling regime (the qubit-fluctuator coupling
strength in the experiment of \cite{NakPasTsaNAT99} is estimated in \cite{GalPRL06}).
In this regime we can insert the result of the one qubit case for
every fluctuator $j$:
\begin{eqnarray*}
\frac{d\tilde{\rho}(t)}{dt} &=&
\sum_j \Gamma^{(j)}_\mathrm{z} \left\{\tilde{s}_\mathrm{z}^{(j)}(t)\tilde{\rho}(t)\tilde{s}_\mathrm{z}^{(j)}(t) -\tilde{\rho}(t)\right\} \\
&& + \Gamma_-^{(j)}\left\{\tilde{s}^{(j)}_-(t)\tilde{\rho}(t)\tilde{s}^{(j)}_+(t)
            - [\tilde{s}^{(j)}_+(t)\tilde{s}^{(j)}_-(t)\tilde{\rho}(t)
							+\tilde{\rho}(t)\tilde{s}^{(j)}_+(t)\tilde{s}^{(j)}_-(t) ]/2\right\}\\
&& + \Gamma_+^{(j)}\left\{\tilde{s}^{(j)}_+(t)\tilde{\rho}(t)\tilde{s}^{(j)}_-(t)
             - [\tilde{s}^{(j)}_-(t)\tilde{s}^{(j)}_+(t)\tilde{\rho}(t)
							 +\tilde{\rho}(t)\tilde{s}^{(j)}_-(t)\tilde{s}^{(j)}_+(t) ]/2\right\} ,
\end{eqnarray*}
where the TLF ladder operators are
$\tilde{s}_\pm \equiv (\tilde{s}_\mathrm{x} \pm \rmi \tilde{s}_\mathrm{y})/2$,
and the decoherence rates are $\Gamma_\mathrm{z}^{(j)} = \gamma_\mathrm{z} \cos^2\theta_j/2$,
$\Gamma_-^{(j)} = (\gamma_- + \gamma_+)\sin^2\theta_j/4$ and
$\Gamma_+^{(j)} = \gamma_+ \sin^2\theta_j/4$.
Here $\gamma_\mathrm{z}$, $\gamma_-$ and $\gamma_+$ are the dephasing rate,
the spontaneous emission rate and stimulated emission rate respectively,
which can be calculated by knowing the spectral properties and temperature of the bath.

Now we return to the Schr\"odinger picture.  For consistency the Heisenberg operators need
to be sent back into the Schr\"odinger picture with the same $\hat{H}_{00}$ used above, so
that in fact we just need to remove the $t$ argument everywhere above because all the
oscillating phase factors accrued actually cancel.
Therefore the final result is:
\begin{eqnarray*}
\frac{d\rho(t)}{dt} &=& -\rmi[\hat{H}_{\mathrm{sq}}+\hat{H}_{\mathrm{TLF}}+\hat{V}_{\mathrm{sq-TLF}},\rho(t)]
+\sum_j \Gamma_\mathrm{z}^{(j)}\left[\sz^{(j)}\rho(t)\sz^{(j)}-\rho(t)\right] \\
&& +\Gamma_-^{(j)}\left[\s^{(j)}_-\rho(t)\s^{(j)}_+ - (\s^{(j)}_+\s^{(j)}_-\rho(t)+\rho(t)\s^{(j)}_+\s^{(j)}_-)/2\right] \\
&& +\Gamma^{(j)}_+\left[\s^{(j)}_+\rho(t)\s^{(j)}_- - (\s^{(j)}_-\s^{(j)}_+\rho(t)-\rho(t)\s^{(j)}_-\s^{(j)}_+)/2\right] .
\end{eqnarray*}
This master equation is valid when $\Omega_j \gg \nu_j$ and $\Omega_j \gg \max\{\lambda_{\ell}\}$, $\forall j$.
The first inequality is needed in the interaction picture (\ref{ApproxExp})
and the second inequality is the standard requirement for the Born-Markov approximation.

Within the same framework we can analyze other situations, such as interacting TLFs, and
additional qubits coupled to the TLFs.  The master equation for interacting TLFs is the
same as above, but with the additional Hamiltonian
$\hat{V}_\mathrm{TLF-TLF}=\sum_{j,k}\mu_{j,k}\sigz^{(j)}\sigz^{(k)}$ and the
requirement that $\min\{ \Omega_j, \Omega_k \} \gg \mu_{j,k}$.
Additional qubits can be included under similar conditions.

\vspace{1cm}
\noindent
{\bf References}
\vspace{1cm}
\bibliographystyle{unsrt}


\end{document}